\documentclass[%
 reprint,
 longbibliography,
 onecolumn,
 amsmath,amssymb,
 aps,
 prfluids,
]{revtex4-2}

\usepackage{graphicx}
\usepackage{dcolumn}
\usepackage{bm}
\usepackage[dvipsnames]{xcolor}

\usepackage[colorlinks,allcolors=blue]{hyperref}
\hypersetup{
    colorlinks,%
    citecolor=blue,%
    filecolor=blue,%
    linkcolor=blue,%
    urlcolor=blue
}

\begin{document}

\title[Non-Boussinesq convection at low Prandtl numbers relevant to the Sun]{Non-Boussinesq convection at low Prandtl numbers relevant to the Sun}

\author{Ambrish Pandey$^{1}$, 
J\"org Schumacher$^{2,3}$, and 
Katepalli R. Sreenivasan$^{1,3,4}$
}
\email{katepalli.sreenivasan@nyu.edu}

\affiliation{$^1$Center for Space Science, New York University Abu Dhabi, Abu Dhabi 129188, United Arab Emirates \\
$^2$Institut f\"ur Thermo- und Fluiddynamik, Technische Universit\"at Ilmenau, Postfach 100565, D-98684 Ilmenau, Germany \\
$^3$Tandon School of Engineering, New York University, New York, NY 11201, USA \\
$^4$Department of Physics and Courant Institute of Mathematical Sciences, New York University, New York, NY 11201, USA}

\date{\today}

\begin{abstract}
Convection in the Sun occurs at Rayleigh numbers, $\mathrm{Ra}$, as high as $10^{22}$, molecular Prandtl numbers, $\mathrm{Pr}$, as low as $10^{-6}$, under conditions that are far from satisfying the Oberbeck-Boussinesq (OB) idealization. The effects of these extreme circumstances on turbulent heat transport are unknown, and no comparable conditions exist on Earth. Our goal is to understand how these effects scale (since we cannot yet replicate the Sun's conditions faithfully). We study thermal convection by using direct numerical simulations, and determine the variation with respect to $\mathrm{Pr}$, to values as low as $10^{-4}$, of the turbulent Prandtl number, $\mathrm{Pr}_t$, which is the ratio of turbulent viscosity to thermal diffusivity. The simulations are primarily two-dimensional but we draw upon some three-dimensional results as well. We focus on non-Oberbeck-Boussinesq (NOB) conditions of a certain type, but also study OB convection for comparison. The OB simulations are performed in a rectangular box of aspect ratio 2 by varying $\mathrm{Pr}$ from $O(10)$ to $10^{-4}$ at fixed Grashof number $\mathrm{Gr} \equiv \mathrm{Ra/Pr} = 10^9$. The NOB simulations are done in the same box by letting only the thermal diffusivity depend on the temperature. Here, the Rayleigh number is fixed at the top boundary while the mean $\mathrm{Pr}$ varies in the bulk from 0.07 to $5 \times 10^{-4}$. The three-dimensional simulations are performed in a box of aspect ratio 25 at a fixed Rayleigh number of $10^5$, and $0.005 \leq \mathrm{Pr} \leq 7$. The principal finding is that $\mathrm{Pr}_t$ increases with decreasing $\mathrm{Pr}$ in both OB and NOB convection: $\mathrm{Pr}_t \sim \mathrm{Pr}^{-0.3}$ for OB convection and $\mathrm{Pr}_t \sim \mathrm{Pr}^{-1}$ for the NOB case. The $\mathrm{Pr}_t$-dependence for the NOB case especially suggests that convective flows in the astrophysical settings behave effectively as in high-Prandtl-number turbulence. 
\end{abstract}

\keywords{Eddy viscosity; eddy diffusivity; turbulent Prandtl number}
\maketitle

\section{Introduction} 
\label{sec:intro}

\subsection{Background}
The Sun has been subject to astronomical observations for many centuries, but a focused study of its internal dynamics began with Johann Fabricius who published sunspot observations in 1611, predating Galileo Galilei's observations in his {\em Lettere solari} by about 2 years. In terms of sustained study in fluid dynamical entities such as waves, oscillations and granulations, as well as instabilities, turbulence and convection, the Sun is a more recent study. The fluid dynamical phenomena that occur in the outer 30\% of the Sun’s radius should be of interest to the readers of this journal and to the members of DFD. It was in this spirit that the Otto Laporte Lecture of the Fluid Dynamics Prize was delivered by the last author of this paper. 

A central theme of that lecture was that the Sun sustains many organized structures though the Rayleigh number of convection is very high, perhaps as high as $10^{22}$, suggesting a domineering role for turbulence. Some important aspects of these organized activities are the differential rotation (i.e., the equator of the Sun rotating faster than the poles), meridional circulation (organized flow from the equator to the poles near the surface and reversal near the bottom of the convection zone), Rossby waves (hydrodynamic, thermal and MHD), granules, supergranules, giant cells, and so forth. The most famous of them are the sunspots that slowly drift towards the equator and whose temporal variation has a period on the order of 22 years (with chaotic variations superimposed on them). How does the Sun, which is a highly turbulent body as suggested by the immense Rayleigh number (and the associated Reynolds number), sustain such highly organized dynamics? This is an important question. 

It is clear that at the heart of the answer lie properties like the enormous stratification with depth of the physical properties of the gas constituting the Sun, its large-scale rotation, the generation of the magnetic field and its interaction with convective fluid dynamics, and the extremely small molecular Prandtl number of the fluid, etc. Time is not yet ripe for a comprehensive and self-contained account of the fluid dynamics of the Sun, a subject to which additions are being made currently at a rapid rate (see e.g. \cite{Schumacher:RMP2020} for a recent review). For this reason it seemed better to narrow the scope of this paper to a specific aspect and provide a detailed account of it, rather than cover the entire range of topics covered in the lecture. 

This specific aspect chosen here is the effect of very small molecular Prandtl number $\mathrm{Pr}$ on the turbulent heat transport. The Prandtl number in the Sun, being of the order of $10^{-6}$, has no analogue in any terrestrial conditions and cannot be replicated in the laboratory, so one has to resort to numerical solutions of governing equations. Further, solar convection does not take place under the Oberbeck-Boussinesq (OB) or near-OB conditions, so one has to build one’s intuition on numerical studies at very low Prandtl numbers under non-Oberbeck-Boussinesq (NOB) conditions. This is the purpose of the rest of the paper, under the joint responsibility of all three authors. For comparisons, we study convection under OB conditions as well. The majority of the simulations are two-dimensional but we also cover three-dimensional convection in a modest parameter space.

\subsection{Specific context}

Turbulent transport mixes substances efficiently and obliterates mean gradients~\citep{Sreenivasan:PNAS2019}. The effective (turbulent) diffusion coefficients in a turbulent flow are far larger than their molecular counterparts and depend on flow properties. Many turbulent flows in nature, including the Sun, are driven by thermal convection~\citep{Hanasoge:ARFM2016, Schumacher:RMP2020}. Rayleigh-B\'enard convection (RBC) is a paradigm for studying the properties of such flows~\citep{Sreenivasan:1998, Ahlers:RMP2009, Chilla:EPJE2012, Verma:NJP2017, Verma:book2018}. In RBC, a horizontal fluid layer is heated from below and cooled from above, and the convective flow properties are governed by the Rayleigh number $\mathrm{Ra}$, the Prandtl number $\mathrm{Pr}$, and the aspect ratio $\Gamma$. The Rayleigh number is a measure of the strength of the driving due to temperature differences compared to friction and diffusive forces due to molecular action. The molecular Prandtl number is given by $\mathrm{Pr} = \nu/\kappa$, the ratio of the kinematic viscosity $\nu$ and the thermal diffusivity $\kappa$ of the fluid. Sometimes it is more convenient, and appropriate, to consider the Grashof number $\mathrm{Gr = Ra/Pr}$. The aspect ratio $\Gamma$ is the ratio of the horizontal to the vertical extents of the domain. The molecular Prandtl number spans a wide range reaching from $\mathrm{Pr} \sim 10^{-6}$ in the Sun's convection region~\citep{Schumacher:RMP2020} to $\mathrm{Pr} \sim 10^{24}$ in the Earth's mantle~\citep{Schubert:book2001}. 

In high-Reynolds-number flows, the turbulent viscosity $\nu_t$ and the turbulent thermal diffusivity $\kappa_t$ are vastly different from their molecular counterparts. The turbulent Prandtl number,  which is given by
\begin{equation}
\mathrm{Pr}_t \equiv \frac{\nu_t}{\kappa_t}\,,
\end{equation}
is a key parameter in various engineering and geo-turbulence models~\citep{Yakhot:IJHMT1987, Otic:NSE2007, Li:AR2019}. For example, $\mathrm{Pr}_t$ is needed to model the convective turbulent heat transport in liquid metal reactors for nuclear energy generation~\citep{Bricteux:NSE2012}. It is essential to know the dependence of $\mathrm{Pr}_t$ on $\mathrm{Pr}$, this being the primary objective of the current work.

According to Reynolds analogy, the eddies which are responsible for the turbulent transport of momentum are also responsible for transporting heat. This yields $\mathrm{Pr}_t \approx 1$. This analogy has been observed to hold reasonably well for convection in air, water, and for $\mathrm{Pr} \geq 0.7$~\citep{Bricteux:NSE2012, Abe:IJHMT2019, Li:AR2019}. However, for convection in liquid metals with $\mathrm{Pr} \ll 1$, $\mathrm{Pr}_t$ values larger than unity have been found ~\citep{Reynolds:IJHMT1975, Jischa:IJHMT1979, Bricteux:NSE2012}. \citet{Abe:IJHMT2019} studied the behavior of $\mathrm{Pr}_t$ in the near-wall region as well as in the central region using direct numerical simulations (DNS) of a channel flow and found that $\mathrm{Pr}_t$ is consistently higher for $\mathrm{Pr} = 0.025$ than for $\mathrm{Pr} = 0.7$. \citet{Bricteux:NSE2012} studied a low-$\mathrm{Pr}$ flow through a uniformly heated channel and observed that $\mathrm{Pr}_t \approx 2$ for $\mathrm{Pr} = 0.01$. Recently, \citet{Tai:PRF2021} studied RBC in a cylindrical cell with $\Gamma = 1$ and observed that $\mathrm{Pr}_t$ within the thermal boundary layer (BL) increased with decreasing $\mathrm{Pr}$.

As $\mathrm{Pr}$ is an inherent property of the fluid, a significant variation of $\mathrm{Pr}$ in experimental studies of convection is possible only when the fluid is changed. Moreover, the lowest $\mathrm{Pr}$ values that can be achieved in a controlled laboratory experiment are of the order of 0.005 for liquid sodium~\citep{Horanyi:IJHMT1999}, more than three orders of magnitude higher than that in the Sun. DNS of RBC, where the governing equations are integrated numerically by resolving all relevant scales, is thus the only available tool for exploring the governing parameters. While the Prandtl number can be varied relatively easily in DNS, in studying very-low-$\mathrm{Pr}$ and very-high-$\mathrm{Ra}$ convection, the challenge of resolving very fine length and time scales limits our scope~\citep{Stevens:JFM2010, Schumacher:PNAS2015, Pandey:POF2016, Scheel:PRF2017, Pandey:Nature2018, Iyer:PNAS2020}. 

Two-dimensional (2D) models of convection are thus commonly utilized to explore parameter dependencies. Such idealized model still provides useful insights on the convective flow properties. For instance, 2D convection has been used to study the properties of convective flow reversals~\citep{Sugiyama:PRL2010, Chandra:PRL2013, Podvin:JFM2015, Pandey:PRE2018}; transition to the so-called ultimate regime of convection~\citep{Zhu:PRL2018}; logarithmic temperature profiles~\citep{Poel:PRL2015}; and the boundary layer (BL) structure~\citep{Zhou:POF2011, Pandey:JFM2021}. Most DNS studies employ the Oberbeck-Boussinesq (OB) model of convection, for which the molecular transport coefficients of the fluid are assumed to be constant throughout the flow~\citep{Chandrasekhar:book, Verma:book2018, Schumacher:RMP2020}. Here, we first study the variation of $\mathrm{Pr}_t$ by performing DNS of 2D-OB convection by varying $\mathrm{Pr}$ over five orders of magnitude, but also compute $\mathrm{Pr}_t$ in a horizontally-extended convection in 3D square box. The relevant 3D data, taken from \citet{Pandey:Nature2018}, correspond to four DNS runs for a constant Rayleigh number $\mathrm{Ra}= 10^5$, with $\mathrm{Pr}$ varying from 0.005 to 7.  The simulations were performed in a rectangular box of dimensions $L_x:L_y:H= 25 : 25 : 1$, where the velocity field satisfies the no-slip boundary condition on all the boundaries. For the temperature field, isothermal and adiabatic conditions are used, respectively, on the horizontal and vertical walls. Thus, this set-up differs from  the corresponding 2D cases, where periodic boundary conditions are employed on the sidewalls.   

Variations of temperature, pressure, and density are assumed to be weak in OB convection in order to safely neglect the variations of the molecular transport properties of the fluid. Thus, it is clear that the OB model is inapplicable in many applications, such as solar, stellar, or even planetary interiors. If one includes these complexities, the resulting NOB convection model becomes very challenging to study~\citep{Zhang:POF1997, Ahlers:JFM2006, Sameen:PhSc2008, Sameen:EPL2009, Sugiyama:JFM2009, Horn:JFM2013, Schumacher:RMP2020, Tritton:book1977}. Recently, in the spirit of ~\citet{Sameen:PhSc2008, Sameen:EPL2009}, we followed a different path and studied a simpler NOB model in a horizontally extended 3D domain. NOB behavior was created by prescribing thermal diffusivity to depend on the temperature, while the other parameters were kept constant~\citep{Getling:PLA2018, Pandey:APJ2021}. We found in ~\citep{Pandey:APJ2021} that many properties of this simple NOB flow are similar to those observed in fully compressible and inelastic models of convection, which are usually utilized to study convection in solar and stellar interiors~\citep{Hanasoge:ARFM2016, Schumacher:RMP2020}. In this paper, we employ the same NOB model in two and three dimensions to study the relation between $\mathrm{Pr}_t$ and $\mathrm{Pr}$. 

The broad finding of this paper is that the turbulent Prandtl number increases with decreasing $\mathrm{Pr}$, the precise rate depending on whether the flow is OB or NOB. To a first approximation, the dimensionality of the flow, the precise sidewall boundary conditions, or the precise values of Rayleigh and Grashof numbers do not matter significantly for the observed variation.  

The paper is organized as follows. In Sec.~\ref{sec:numerical}, we describe the flow domain, the numerical method, and the parameters explored, in both OB and NOB convection. In Sec.~\ref{sec:OB}, we present the flow structure and behavior of $\mathrm{Pr}_t$ in OB convection. The scaling of integral quantities, flow structures and the variation of $\mathrm{Pr}_t$ in NOB convection are presented in Sec.~\ref{sec:NOB}. Finally, the important findings from the current study are summarized in Sec.~\ref{sec:conclu}, where we return to implications for the Sun.

\section{Governing equations and numerical details}
\label{sec:numerical}

\subsection{Two-dimensional Oberbeck-Boussinesq convection}
\label{sec:nume_OB}

We study the OB model of convection in a 2D domain by performing DNS from $\mathrm{Pr} = 10^{-4}$ to $12.73$ at a constant Grashof number $\mathrm{Gr} = 10^9$. The relevant non-dimensional governing equations are~\citep{Schumacher:PNAS2015}
\begin{eqnarray}
\frac{\partial {\bm u}}{\partial t} + {\bm u} \cdot \nabla {\bm u} & = & -\nabla p + T \hat{z} +  \frac{1}{\sqrt{\mathrm{Gr}}}   \nabla^2 {\bm u}, \label{eq:u} \\ 
\frac{\partial T}{\partial t} + {\bm u} \cdot \nabla T & = &  \frac{1}{\sqrt{\mathrm{Gr}} \mathrm{Pr}} \nabla^2 T, \label{eq:T} \\ 
\nabla \cdot {\bm u} & = & 0, \label{eq:m}
\end{eqnarray}
where ${\bm u} = (u_x,u_z), p$, and $T$ are respectively the velocity, pressure, and temperature fields defined on a rectangular domain of horizontal dimension $L = \Gamma H$ and vertical height $H$. The Rayleigh number is defined as $\mathrm{Ra} = \gamma g \Delta T H^3/(\nu \kappa)$, $\gamma$ being the (isobaric) thermal expansion coefficient of the fluid, $g$ the acceleration due to gravity, and $\Delta T$ the temperature difference between bottom and top plates. These equations are non-dimensionalized using $H$, the free-fall velocity $u_f = \sqrt{\gamma g \Delta T H}$, the free-fall time $t_f = H/u_f$, and $\Delta T$ as length, velocity, time, and temperature scales, respectively.

We perform the DNS of OB convection by integrating Eqs.~(\ref{eq:u})--(\ref{eq:m}) using a spectral element solver {\sc Nek5000}~\citep{Fischer:JCP1997}. The flow domain is discretized into a finite number of elements ($N_e$) and the turbulence fields within each element are expanded using $N^\mathrm{th}$ order Lagrangian interpolation polynomials, resulting in $N_e N^2$ mesh cells in the flow. The top and bottom plates satisfy isothermal and no-slip boundary conditions, whereas the sidewalls are periodic. The simulations are initiated from the conduction state with random perturbations and the analyses are performed after the initial transients have decayed. To sufficiently resolve the boundary layers near the horizontal plates, we place a larger number of mesh cells in those regions than in the bulk. We have verified that the flows are spatially well-resolved according to criteria summarized in~\citet{Scheel:NJP2013} and  \citet{Pandey:JFM2021}. 

Convective flows at fixed $\mathrm{Ra}$ become increasingly turbulent as $\mathrm{Pr}$ is lowered~\citep{Schumacher:PNAS2015, Pandey:POF2016, Pandey:Nature2018, Pandey:EPL2021}. This increases the computational cost for exploring low-$\mathrm{Pr}$ convection even at moderate Rayleigh numbers because the finest length and time scales, namely the Kolmogorov scales (or smaller), need to be properly resolved. The study of convection with varying $\mathrm{Pr}$ for a given $\mathrm{Ra}$ becomes extremely challenging when $\mathrm{Pr} \ll 1$~\citep{Scheel:PRF2017}. So we study convective flows at a constant Grashof number $\mathrm{Gr} = 10^9$ from $\mathrm{Pr} = O(10)$ down to $\mathrm{Pr} = 10^{-4}$. As mentioned already, the Rayleigh number is also simultaneously varied to keep $\mathrm{Gr}$ a constant. Note that the momentum Eq.~\eqref{eq:u} remains the same for flows at a constant $\mathrm{Gr}$, and the Prandtl number appears explicitly only in the temperature equation. However, as the momentum and temperature evolution equations are coupled, the momentum equation also feels the change in $\mathrm{Pr}$ via the temperature field~\citep{Schumacher:PNAS2015}. Table~\ref{table:details_OB} summarizes important simulation parameters, and shows the challenges in studying very low-$\mathrm{Pr}$ convection: note that the same spatial resolution is required for OB1 and OB7 simulations, even though $\mathrm{Ra}$ is smaller by five orders of magnitude in the latter.
\begin{table}
\begin{center}
\caption{Parameters of the OB simulations in a $\Gamma = 2$ box with a fixed Grashof number $\mathrm{Gr} = 10^9$. The total number of spectral elements is $N_e = 46,000$. Here, $N$ is the order of the Lagrangian interpolation polynomials; $\mathrm{Nu}$, $\mathrm{Nu}_{\varepsilon_T}$, and $\mathrm{Nu}_{\varepsilon_u}$ are the Nusselt numbers computed using Eqs.~(\ref{eq:Nu_uzT}), (\ref{eq:Nu_epst}) and~(\ref{eq:Nu_epsu}), respectively; $\mathrm{Re}$ is Reynolds number computed using the root-mean-square velocity. The ``error bars" in $\mathrm{Nu}$ and $\mathrm{Re}$ are the differences between the mean values from the first and second halves of the datasets.}
\begin{ruledtabular}
{\begin{tabular}{cccccccc}
Run & $\mathrm{Pr}$ & $\mathrm{Ra}$ & $N$ & $\mathrm{Nu}$ & $\mathrm{Nu}_{\varepsilon_T}$ & $\mathrm{Nu}_{\varepsilon_u}$  & $\mathrm{Re}$ \\
\hline
OB1 & 12.73 & $1.273 \times 10^{10}$ & 7 &  $104.2 \pm 1.3$ & 103.7 & 103.7 & $6103 \pm 844$ \\
OB2 & 0.7 & $7 \times 10^8$ & 3 &  $43.0 \pm 1.6$ & 43.3 & 43.2 & $16429 \pm 4095$ \\
OB3 & 0.1 & $1 \times 10^8$ & 5 &  $23.7 \pm 0.6$ & 23.4 & 23.4 & $28605 \pm 5200$ \\
OB4 & 0.02546 & $2.546 \times 10^7$ & 5 & $14.8 \pm 0.3$ & 14.8 & 14.9 & $41477 \pm 11350$ \\
OB5 & $0.005$ & $5 \times 10^6$ & 7 &  $8.43 \pm 0.01$ & 8.41 & 8.29 & $61771 \pm 14510$ \\
OB6 & $0.001$ & $1 \times 10^6$ & 7 &  $4.52 \pm 0.17$ & 4.53 & 4.47 & $98920 \pm 16465$ \\
OB7 & $0.0001$ & $1 \times 10^5$ & 7 &  $1.69 \pm 0.05$ & 1.69 & 1.73 & $147811 \pm 21082$ \\
\end{tabular}}
\end{ruledtabular}
\label{table:details_OB}
\end{center}
\end{table}

\subsection{Two-dimensional non-Oberbeck-Boussinesq convection}
\label{sec:nume_NOB}

The relevant incompressible ($\nabla \cdot {\bm u} = 0$) nondimensional governing equations are
\begin{eqnarray}
\frac{\partial {\bm u}}{\partial t} + {\bm u} \cdot \nabla {\bm u} & = & -\nabla p + T \hat{z} +\frac{1}{\sqrt{\mathrm{Gr}}} \, \nabla^2 {\bm u}, \label{eq:u_n} \\
\frac{\partial T}{\partial t} + {\bm u} \cdot \nabla T & = & \nabla \cdot \left[ \frac{f(T)}{\sqrt{\mathrm{Gr}} \, \mathrm{Pr}_{\mathrm{top}} } \nabla T \right].
 \label{eq:T_n}
\end{eqnarray}
The Rayleigh and Prandtl numbers at the top boundary are defined as $\mathrm{Ra}_\mathrm{top} = \gamma g \Delta T H^3/(\nu \kappa_\mathrm{top})$ and $\mathrm{Pr}_\mathrm{top} = \nu/\kappa_\mathrm{top}$. The NOB simulations are also performed in the same 2D box of $\Gamma = 2$ by integrating Eqs.~(\ref{eq:m})--(\ref{eq:T_n}) using {\sc Nek5000}~\citep{Fischer:JCP1997}. 
Following \citet{Pandey:APJ2021}, we use two different functional forms of $\kappa(T)$, which are given by
\begin{eqnarray}
\kappa_1(T) & = & \kappa_\mathrm{top}(1+49T+450T^6), \label{eq:k1} \\
\kappa_2(T) & = & \kappa_\mathrm{top}(1+149T+350T^3). \label{eq:k2}
\end{eqnarray}
Thus, $\kappa(T)$ increases towards the heated bottom plate, where the diffusivity in both cases is $\kappa_\mathrm{bot} = 500 \kappa_\mathrm{top}$. The parameters are $\mathrm{Pr}_\mathrm{top} = 12.73$ and $\mathrm{Ra}_\mathrm{top} = 1.708 \times 10^8$, corresponding to $\mathrm{Pr}_\mathrm{bot} = 0.025$ and $\mathrm{Ra}_\mathrm{bot} = 3.4 \times 10^5$. The Grashof number is $\mathrm{Gr} = \mathrm{Ra}_\mathrm{bot}/\mathrm{Pr}_\mathrm{bot} = 1.34 \times 10^7$. We have performed four more simulations with the diffusivity variation specified by $\kappa_2(T)$ for $\mathrm{Pr}_\mathrm{top} = 1.0, 0.5, 0.25$, and 0.1, while $\mathrm{Ra}_\mathrm{top} = 1.708 \times 10^8$. Important parameters of the NOB simulations are summarized in Table~\ref{table:details_NOB}.

The strongly varying temperature field in the vicinity of the top plate requires a finer local spatial resolution which is adjustable in a spectral element method. Therefore, we designed an asymmetric mesh containing larger number of grid points near the top plate than near the bottom plate, and verified that all the NOB simulations were adequately resolved as discussed in~\citet{Pandey:APJ2021}.
\begin{table}
\begin{center}
\caption{Parameters of the NOB simulations for $\mathrm{Ra}_\mathrm{top} = 1.708 \times 10^8$. The simulation domain is divided into 9900 spectral elements. Here, $\kappa(T)$ is the functional form of the temperature-dependent thermal diffusivity specified by Eqs.~(\ref{eq:k1}) and~(\ref{eq:k2}); $\mathrm{Pr}_\mathrm{top}$ is the Prandtl number specified at the top plate; $\mathrm{Gr}$ is the Grashof number; $\langle \mathrm{Pr} \rangle$ is the globally-averaged Prandtl number; $N$ is the order of the Lagrangian interpolation polynomials. The error bars in $\mathrm{Nu}$ and $\mathrm{Re}$ are computed as in Table~\ref{table:details_OB}.}
\begin{ruledtabular}
{\begin{tabular}{cccccccccc}
Run & $\kappa(T)$ & $\mathrm{Pr}_\mathrm{top}$ & $\mathrm{Gr}$ & $\langle \mathrm{Pr} \rangle$ & $N$ & $\mathrm{Nu}$ & $\mathrm{Nu}_{\varepsilon_T}$ & $\mathrm{Nu}_{\varepsilon_u}$  & $\mathrm{Re}$ \\
\hline
NOB1 & $\kappa_1(T)$ & $12.73$ & $1.342 \times 10^7$ & $7.2 \times 10^{-2}$ & 5 &  $6.57 \pm 0.01$ & 6.57 & 6.57 & $2048 \pm 1$ \\
NOB2 & $\kappa_2(T)$  & $12.73$ & $1.342 \times 10^7$ & $5.3 \times 10^{-2}$ & 5 &  $6.23 \pm 0.02$ & 6.24 & 6.23 & $2574 \pm 3$ \\
NOB3 & $\kappa_2(T)$ & $1.00$ & $1.708 \times 10^8$  & $4.4 \times 10^{-3}$ & 7 & $4.72 \pm 0.07$ & 4.72 & 4.67 & $20350 \pm 176$ \\
NOB4 & $\kappa_2(T)$ & $0.50$ & $3.416 \times 10^8$   & $2.2 \times 10^{-3}$ & 11 & $4.19 \pm 0.24$ & 4.19 & 3.98 & $34740 \pm 855$ \\
NOB5 & $\kappa_2(T)$ & $0.25$ & $6.832 \times 10^8$   & $1.1 \times 10^{-3}$ & 13 & $3.73 \pm 0.04$ & 3.77 & 3.60 & $59634 \pm 229$ \\
NOB6 & $\kappa_2(T)$ & $0.10$ & $1.708 \times 10^9$   & $4.7 \times 10^{-4}$  & 13 &  $3.52 \pm 0.06$ & 3.53 & 3.51 & $121752 \pm 84$ \\
\end{tabular}}
\end{ruledtabular}
\label{table:details_NOB}
\end{center}
\end{table}

\begin{figure}
\includegraphics[width=0.85\textwidth]{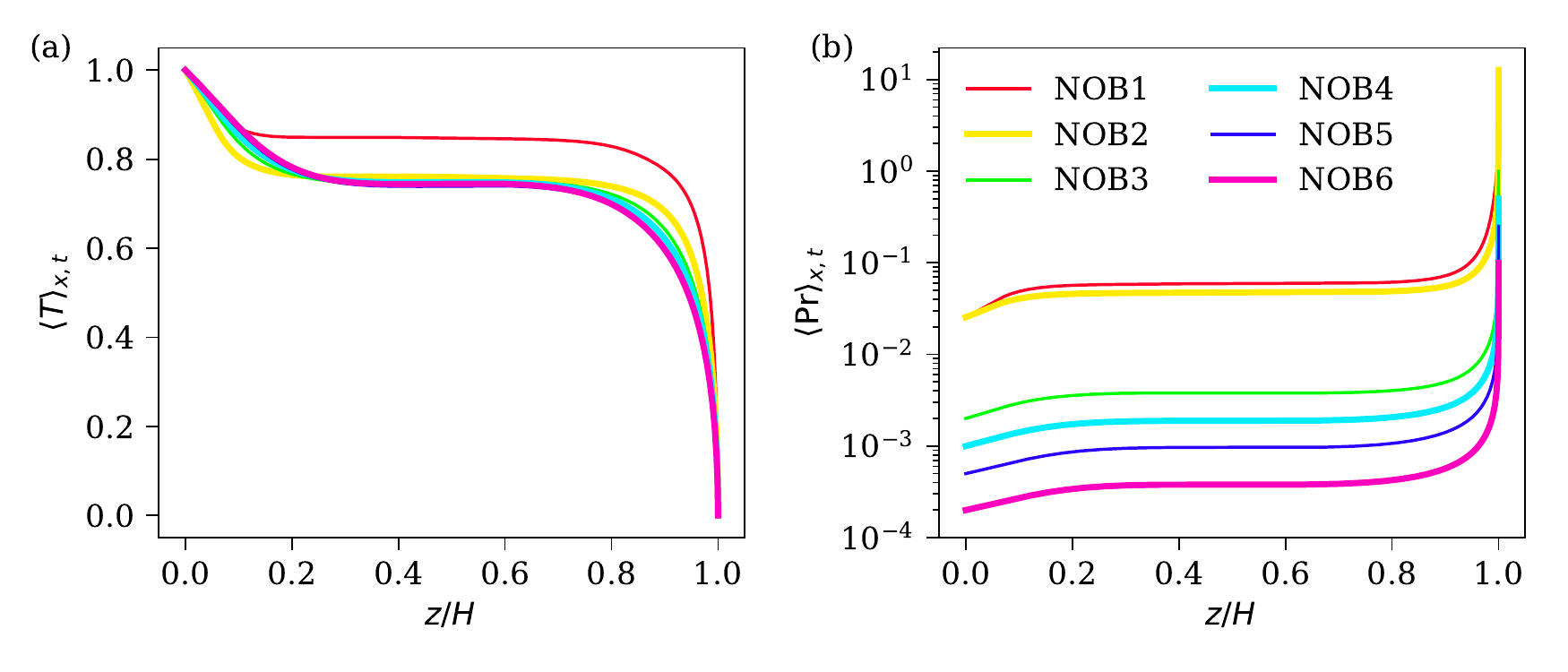}
\caption{Horizontally- and temporally-averaged (a) temperature and (b) Prandtl number profiles for NOB simulations show that the isothermal bulk region shrinks with decreasing $\mathrm{Pr}$. The profiles are asymmetric with respect to the midline at $z=0.5 H$.}
\label{fig:T_Pr_z}
\end{figure}

One consequence of using a temperature-dependent diffusivity is that the temperature profile becomes asymmetric with respect to midplane~\citep{Pandey:APJ2021} (see Fig.~\ref{fig:T_Pr_z}(a)).
We compute the horizontally- and temporally averaged temperature profile $\langle T \rangle_{x,t}$. Figure~\ref{fig:T_Pr_z}(a) shows that $\langle T \rangle_{x,t}$ does not vary appreciably in the bulk of the flow. The profiles are asymmetric with respect to the mid-line, i.e., the temperature drop is larger in the top region than in the bottom region, which is a signature of the NOB nature of the flow~\citep{Zhang:POF1997, Horn:JFM2013}.  Figure~\ref{fig:T_Pr_z}(a) also reveals that the stratification in the top region in NOB1 is stronger than that in NOB2, which is due to a weaker variation of $\kappa$ with temperature in the latter. The mean temperature in NOB2 is smaller than that in NOB1, which indicates that the departure from OB conditions becomes weaker with decreasing polynomial order of $\kappa(T)$. We have verified this result by performing an additional simulation with the thermal diffusivity varying linearly with $T$, i.e., $\kappa(T) = \kappa_\mathrm{top}(1+499 T)$, but do not present the results to avoid clutter.

We compute the depth-variation of the horizontally-averaged Prandtl number, $\langle \mathrm{Pr} \rangle_{x,t} = \nu/\langle \kappa(T) \rangle_{x,t}$, and plot it in Fig.~\ref{fig:T_Pr_z}(b). The figure shows that $\mathrm{Pr}$ drops sharply from its value at the top plate and does not change appreciably in the bulk region. It shows that $\mathrm{Pr}$ in the bulk is nearly two orders of magnitude lower than that at the top plate. We compute the mean $\mathrm{Pr}$ of the flow as $\langle \mathrm{Pr} \rangle = \nu/\langle \kappa \rangle_{A,t}$, where $\langle \cdot \rangle_{A,t}$ denotes the averaging over the entire simulation domain and time, and list the values of $\langle \mathrm{Pr} \rangle$ in Table~\ref{table:details_NOB} and find that $\langle \mathrm{Pr} \rangle < 0.1$ for all the simulations. Thus, the flow properties of our NOB simulations are similar to those observed in low-$\mathrm{Pr}$ convection. 

\subsection{Supporting results from three-dimensional Oberbeck-Boussinesq convection}

To assess the robustness of 2D results, we study 3D convection in a square box, utilizing the database from \citet{Pandey:Nature2018}, where the properties of turbulent convective superstructures were explored for $\Gamma = 25$. We select four cases for a fixed Rayleigh number $\mathrm{Ra} = 10^5$, where $\mathrm{Pr}$ is varied from 0.005 to 7. These simulations also use the {\sc Nek5000} spectral element solver. The parameters of these four simulations are listed in Table~\ref{table:details_OB_3D}.
\begin{table}
\begin{center}
\caption{Parameters of the OB simulations in a $\Gamma = 25$ 3D rectangular box with a fixed Rayleigh number $\mathrm{Ra} = 10^5$ (taken from \citet{Pandey:Nature2018}). OB1 is not a turbulent flow for these flow conditions and will not be considered further.}
\begin{ruledtabular}
{\begin{tabular}{cccccccc}
Run & $\mathrm{Pr}$ & $N_e$ & $N$ & $\mathrm{Nu}$ & $\mathrm{Nu}_{\varepsilon_T}$ & $\mathrm{Nu}_{\varepsilon_u}$  & $\mathrm{Re}$ \\
\hline
3D OB1 & 7.0 & 1352000 & 5 & $4.1 \pm 0.01$ & 4.1 & 4.1 & $11 \pm 0.03$ \\
3D OB2 & 0.7 & 1352000 & 5 & $4.3 \pm 0.02$ & 4.3 & 4.3 & $92 \pm 0.4$ \\
3D OB3 & 0.021 & 2367488 & 7 & $2.6 \pm 0.01$ & 2.6 & 2.6 & $1120 \pm 8$ \\
3D OB4 & 0.005 & 2367488 & 11 & $1.9 \pm 0.01$ & 1.9 & 1.9 & $2491 \pm 20$ \\
\end{tabular}}
\end{ruledtabular}
\label{table:details_OB_3D}
\end{center}
\end{table}
     
\section{Turbulent Prandtl number in Oberbeck-Boussinesq  convection} 
\label{sec:OB}

\subsection{Flow structure and global transport of heat and momentum}
\label{sec:str_OB}

\begin{figure}
\includegraphics[width=0.9\textwidth]{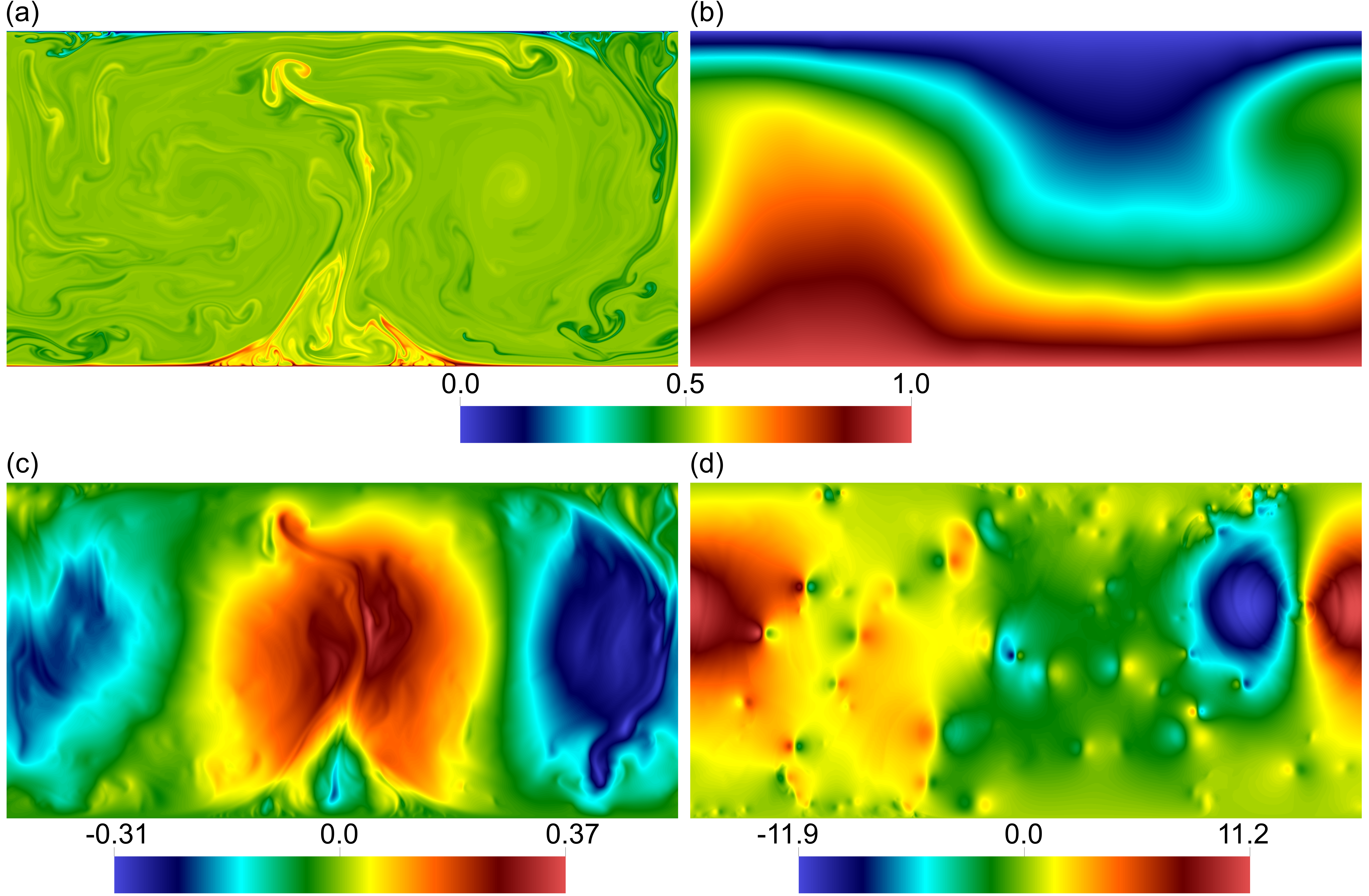}
\caption{Instantaneous temperature (a, b) and vertical velocity (c, d) fields for OB1 (a, c) and OB7 (b, d) simulations. The temperature variation is much smoother and the thermal structures are coarser in (b) than in (a), whereas the velocity field in (d) is much more patchy and intermittent than in (c).}
\label{fig:T_uz_OB}
\end{figure}

Low-$\mathrm{Pr}$ convection is characterized by larger contrast in the scales of the velocity and temperature fields~\citep{Scheel:PRF2017, Pandey:Nature2018, Schumacher:RMP2020}. The temperature field is highly diffusive due to a shorter thermal diffusion time scale $H^2/\kappa$ compared to the momentum diffusion time scale $H^2/\nu$ in low-$\mathrm{Pr}$ convection. The finest scales of the velocity field are thus much finer compared to those of the temperature field; in contrast, in high-$\mathrm{Pr}$ convection flow, the temperature field exhibits very fine structures~\citep{Silano:JFM2010, Horn:JFM2013, Pandey:PRE2014, Pandey:Pramana2016}. To illustrate this, we show in Fig.~\ref{fig:T_uz_OB} instantaneous snapshots of temperature and the vertical velocity fields for OB1 and OB7. The characteristic scales of the thermal structures are very different in the two flows. The thickness of the thermal structures is similar to the thermal BL width $\delta_T$, which is related to the Nusselt number $\mathrm{Nu}$ as $\delta_T = 0.5H/\mathrm{Nu}$~\citep{Pandey:JFM2021}. The Nusselt number quantifies the turbulent heat transport in a convective flow and is defined as the ratio of the total to the conductive heat transport. We compute $\mathrm{Nu}$ using the simulation data as
\begin{equation}
\mathrm{Nu} = 1 + \sqrt{\mathrm{Ra Pr}} \, \langle u_z T \rangle_{A,t}. \label{eq:Nu_uzT}
\end{equation}
The Nusselt numbers for all the OB simulations are listed in Table~\ref{table:details_OB}. We find $\mathrm{Nu} \approx 104$ for OB1 and 1.7 for OB7, which yield $\delta_T \approx 0.005 H$ and $0.3 H$ for the two cases. Thus, the total heat transport is dominated by the molecular diffusion for OB7 ($\mathrm{Pr} = 10^{-4}$), whereas by turbulent convection for OB1 ($\mathrm{Pr} = 12.73$).

The global heat transport is related to the globally-averaged thermal and viscous dissipation rates as~\citep{Shraiman:PRA1990}
\begin{eqnarray}
\mathrm{Nu}_{\varepsilon_T} & = & \sqrt{\mathrm{Ra Pr}} \, \langle \varepsilon_T \rangle_{A,t} \, , \label{eq:Nu_epst} \\
\mathrm{Nu}_{\varepsilon_u} & = & 1 + \sqrt{\mathrm{Ra Pr}} \, \langle \varepsilon_u \rangle_{A,t} \, , \label{eq:Nu_epsu}
\end{eqnarray}
where $\varepsilon_T$ and $\varepsilon_u$ are the thermal and viscous dissipation rates, respectively, defined by
\begin{eqnarray}
\varepsilon_T({\bm x}) & = & \kappa(T) \left( \frac{\partial T}{\partial x_i} \right)^2, \label{eq:epst} \\
\varepsilon_u({\bm x}) & = & \frac{\nu}{2} \left( \frac{\partial u_i}{\partial x_j} + \frac{\partial u_j}{\partial x_i} \right)^2 \label{eq:epsu},
\end{eqnarray}
with $u_i$ being the $i^\mathrm{th}$-component of the velocity field. The $\mathrm{Nu}_{\varepsilon_T}$ and $\mathrm{Nu}_{\varepsilon_u}$, provided in Table~\ref{table:details_OB}, agree very well with $\mathrm{Nu}$ for all the simulations. This is another indication that the numerical resolution is adequate for all the simulations~\citep{Pandey:JFM2021}. The turbulent momentum transport is quantified using the Reynolds number $\mathrm{Re}$, which is computed as $\mathrm{Re} = \sqrt{\mathrm{Ra/Pr}} \, u_\mathrm{rms}$ with $u_\mathrm{rms} = \langle u_i^2 \rangle_{A,t}^{1/2}$ as the root-mean-square velocity. Table~\ref{table:details_OB} lists $\mathrm{Re}$ for all the simulations and we observe that $\mathrm{Re}$ increases with decreasing $\mathrm{Pr}$. The velocity field $u_z(x,z,t_0)$ for OB7 in Fig.~\ref{fig:T_uz_OB}(d) exhibits fine vortex structures with broader range of length scales compared to OB1, which is smoother. For the three-dimensional case the average $\langle \cdot\rangle_{A,t}$ has to be substituted by a combined volume and time average $\langle\cdot\rangle_{V,t}$ in different Nusselt number definitions. 

\begin{figure}
\includegraphics[width=0.55\textwidth]{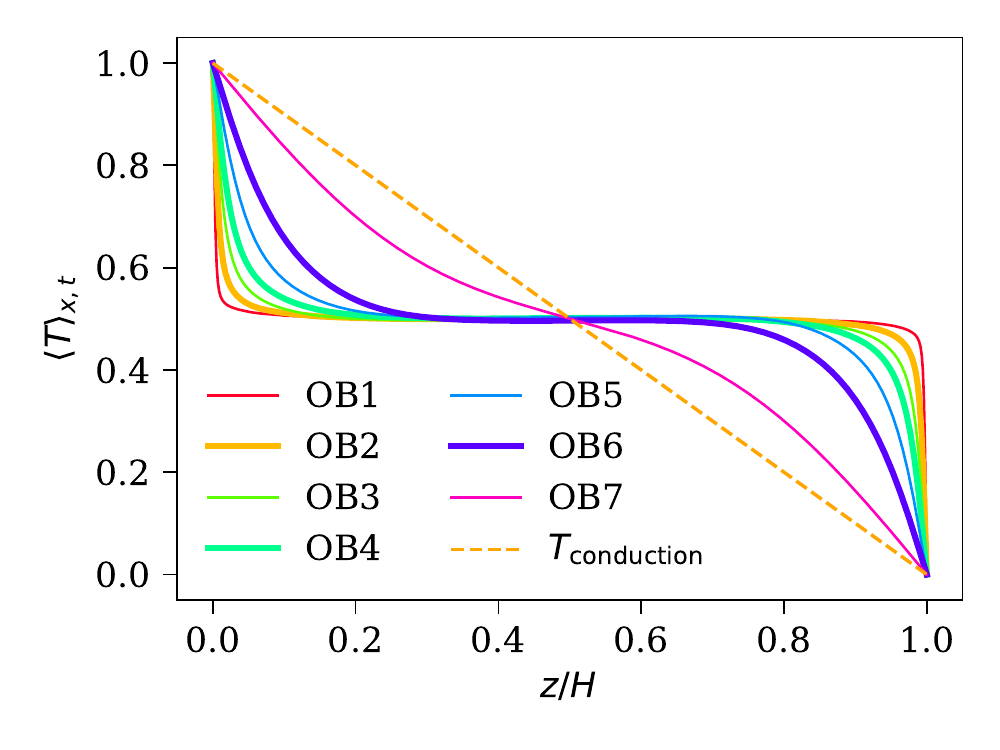}
\caption{Temperature profiles for OB simulations show that the isothermal bulk region shrinks with decreasing $\mathrm{Pr}$ and barely exists for $\mathrm{Pr} = 10^{-4}$. The dashed line corresponds to the linear profile in the conduction state $T_\mathrm{conduction} = 1 - z$ in dimensionless form.}
\label{fig:T_z_OB}
\end{figure}

Figure~\ref{fig:T_uz_OB} also reveals that the bulk of the flow is well-mixed and nearly isothermal for OB1 (panel a), whereas a strong temperature gradient is present in OB7 (panel b). We quantify this by plotting $\langle T \rangle_{x,t}$ as a function of $z$ in Fig.~\ref{fig:T_z_OB}, which shows that, except near the plates within the thermal BLs, $\langle T \rangle_{x,t} \approx 0.5$ in the bulk of the flow for OB1. Thus, the mean temperature in the bulk region is the arithmetic mean of the prescribed values at the top and bottom, which is a characteristic of the OB convection. For OB7, however, the temperature profile departs only slightly from the linear conduction profile, with no well-mixed bulk region in the flow. 

\begin{figure}
\begin{center}
\includegraphics[width=0.8\textwidth]{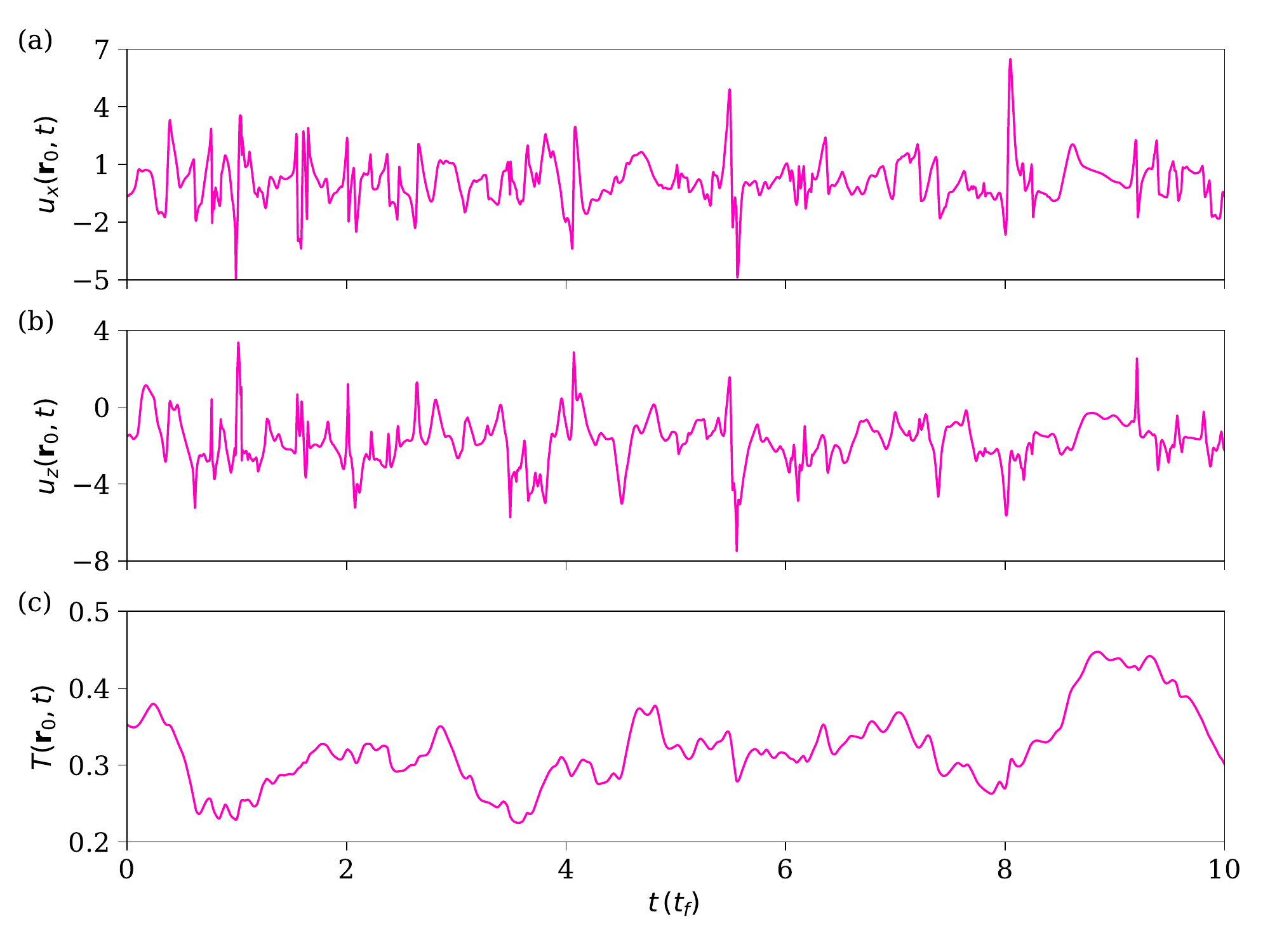}
\caption{Time traces of (a) horizontal velocity, (b) vertical velocity, and (c) temperature taken at the center at ${\bf r}_0 = (L/2, H/2)$ for run OB7 with $Pr = 10^{-4}$ and $Ra = 10^5$. A shorter segment of the entire time trace is shown only to highlight the irregular and stochastic nature of all fields. The signals indicate that the flow is turbulent despite the moderate Rayleigh number, but that the temperature field is coarse due to high diffusivity.}
\label{fig:trace}
\end{center}
\end{figure}

It has been reported that the flow at low Prandtl numbers possesses convection rolls which are similar to `flywheel', with fluid rotating rigidly near the axis of the rolls~\citep{Jones:JFM1976, Clever:JFM1981, Busse:JFM1981}. However, this laminar `flywheel' state has been observed for low Rayleigh numbers which are not far from the onset of convection~\citep{Thual:JFM1992}. For $\mathrm{Pr} = 10^{-4}$ at $\mathrm{Ra} = 10^5$, the flow is not at all laminar, but highly turbulent. This is corroborated by the time trace of velocity and temperature at a fixed position in the flow. We record the time evolution of the velocity and temperature fields at various positions in the flow, and show them in Fig.~\ref{fig:trace} for the center of the domain for OB7. Both the velocity and temperature fields vary turbulently at the center (and elsewhere as well).

\subsection{The turbulent Prandtl number}
\label{sec:Prt_OB}

To estimate the turbulent viscosity $\nu_t$ and the turbulent thermal diffusivity $\kappa_t$, we decompose the velocity and temperature fields into their mean and fluctuating parts,
\begin{eqnarray}
{\bm u({\bm x},t)} & = & {\bm U}({\bm x}) + {\bm u^\prime}({\bm x},t), \\
T({\bm x},t) & = & \Theta({\bm x}) + T^\prime({\bm x},t), 
\end{eqnarray}
where ${\bm U(\bm x)}$ and $\Theta({\bm x})$ are the time-averaged velocity and temperature fields. In the literature, the turbulent viscosity is usually estimated by the flux-gradient method, according to which $\nu_t = - \langle u_x^{\prime} u_z^{\prime} \rangle/(\partial U_x/\partial z)$ and the turbulent thermal diffusivity by $\kappa_t = - \langle u_z^{\prime} T^{\prime} \rangle/(\partial \Theta/\partial z)$~\citep{Bricteux:NSE2012, Emran:JFM2015, Shishkina:PRF2017, Ching:PRR2019, Li:AR2019, Pandey:JFM2021, Tai:PRF2021}. In turbulent convection, however, both $\nu_t$ and $\kappa_t$ computed using this method become undefined at some heights. This is because the mean velocity gradient $\partial U_x/\partial z$ often changes sign due to the absence of a well-defined mean flow. To overcome this ambiguity, we use the $k-\varepsilon$ approach ($k$ will be denoted as $k_u$ in the following), according to which the turbulent diffusivities can be estimated by
\begin{eqnarray}
\nu_t & = & c_\nu k_u^2 /\varepsilon_{u^\prime} \, , \\
\kappa_t & = & c_\kappa k_u k_T /\varepsilon_{T^\prime} \, ,
\end{eqnarray}
where $k_u = \langle{\bm u^{\prime 2}}\rangle/2$ is the turbulent kinetic energy and $k_T = \langle T^{\prime 2}\rangle$ is the thermal variance, and $\varepsilon_{u^\prime}$ and $\varepsilon_{T^\prime}$ are, respectively, the mean turbulent viscous and thermal dissipation rates, computed from equations~(\ref{eq:epst}) and~(\ref{eq:epsu}). In this approach, $\nu_t$ is proportional to the square of the turbulent kinetic energy; this is plausible because stronger turbulent fluctuations produce stronger turbulent mixing, resulting in higher turbulent viscosity~\citep{Davidson:book2004}. In engineering turbulence models, the proportionality constant $c_\nu \approx 0.09$ is most often used~\citep{Davidson:book2004}; it has also been found by renormalization group theory in ~\citet{Yakhot:JSC1986, Yakhot:IJHMT1987}. The coefficient $c_\kappa$ is less well known, though $c_\kappa \approx 0.1$ has been considered in an RBC simulation~\citep{Otic:NSE2007}. Not knowing the prefactors at low molecular Prandtl numbers, we leave the specific values of these coefficients open, as we are interested primarily in the variation of the ratio $\nu_t/\kappa_t$. To fix the ratio $c_\nu/c_\kappa$, we choose it such that the turbulent Prandtl number in our flow agrees with that observed in the literature for $\mathrm{Pr} = 0.7$. We also treat $c_\nu/c_\kappa$ as a constant independent of $\mathrm{Pr}$ and $\mathrm{Ra}$.

\begin{figure}
\includegraphics[width=0.9\textwidth]{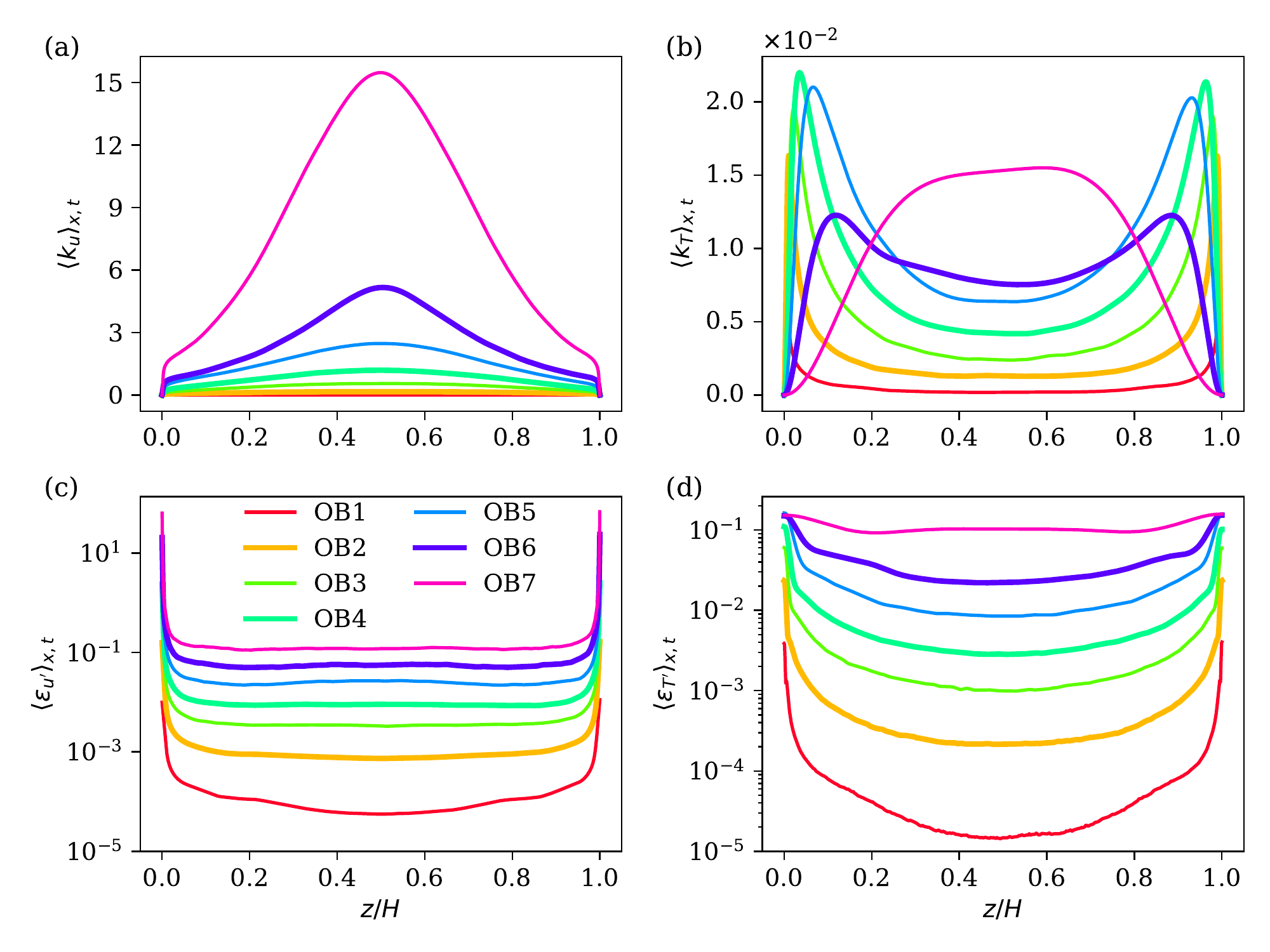}
\caption{Vertical profiles of the (a) turbulent kinetic energy, (b) thermal variance, (c) turbulent viscous dissipation rate, and (d) turbulent thermal dissipation rate for 2D OB simulations. The turbulent Prandtl number $\mathrm{Pr}_t(z)$ is computed using these profiles and is shown in Fig.~\ref{fig:Prt_OB}.}
\label{fig:profiles_comp_OB}
\end{figure}

The vertical profiles $k_u(z), k_T(z), \varepsilon_{u^\prime}(z)$, and $\varepsilon_{T^\prime}(z)$ for all the simulations are shown in Fig.~\ref{fig:profiles_comp_OB}. We note that $k_u$ and $k_T$ vanish at the plates because of the no-slip and isothermal boundary conditions. Figure~\ref{fig:profiles_comp_OB}(a) shows that $k_u$ exhibits a maximum in the center plane at $z \approx H/2$~\citep{Deardorff:JFM1967, Adrian:IJHMT1996}. The turbulent kinetic energy increases with decreasing $\mathrm{Pr}$ (see table~\ref{table:details_OB}). Figure~\ref{fig:profiles_comp_OB}(b) shows that, in contrast to $k_u$, $k_T$ exhibits maxima near the plates and decreases towards the central region. The maxima of $k_T$ occur near the edge of the thermal BL; this makes sense because it is the contrast in the temperatures of the plumes and the ambient fluid that causes these maxima~\citep{Pandey:JFM2021, Pandey:APJ2021}. The peaks in $k_T$ are, however, absent for $\mathrm{Pr} = 10^{-4}$, which indicates the presence of very thick thermal BLs, extending all the way to the center plane, leading to the absence of a well-mixed bulk region, consistent with Fig.~\ref{fig:T_z_OB}. 

Figure~\ref{fig:profiles_comp_OB}(c,d) demonstrate that the turbulent viscous and thermal dissipation rates are largest at the plates and decrease into the bulk region. We also observe that the dissipation rates increase with decreasing $\mathrm{Pr}$; the non-dimensional dissipation rates are computed as $\varepsilon_{u^\prime} \sim (\mathrm{Nu}-1)/(\sqrt{\mathrm{Gr}} \mathrm{Pr})$ and $\varepsilon_{T^\prime} \sim \mathrm{Nu}/(\sqrt{\mathrm{Gr}} \mathrm{Pr})$. Our data show that $\mathrm{Nu}$ increases as $\mathrm{Pr}^{0.35}$ for $\mathrm{Gr} = 10^9$. This yields $\varepsilon_{u^\prime} \sim \mathrm{Pr}^{-0.65}$ and $\varepsilon_{T^\prime} \sim \mathrm{Pr}^{-0.65}$ as $\mathrm{Gr}$ is the same for all the simulations. The increase of the dissipation rates with decreasing $\mathrm{Pr}$ for a constant $\mathrm{Gr}$ is also consistent with~\citet{Scheel:JFM2016}.

\begin{figure}
\includegraphics[width=1\textwidth]{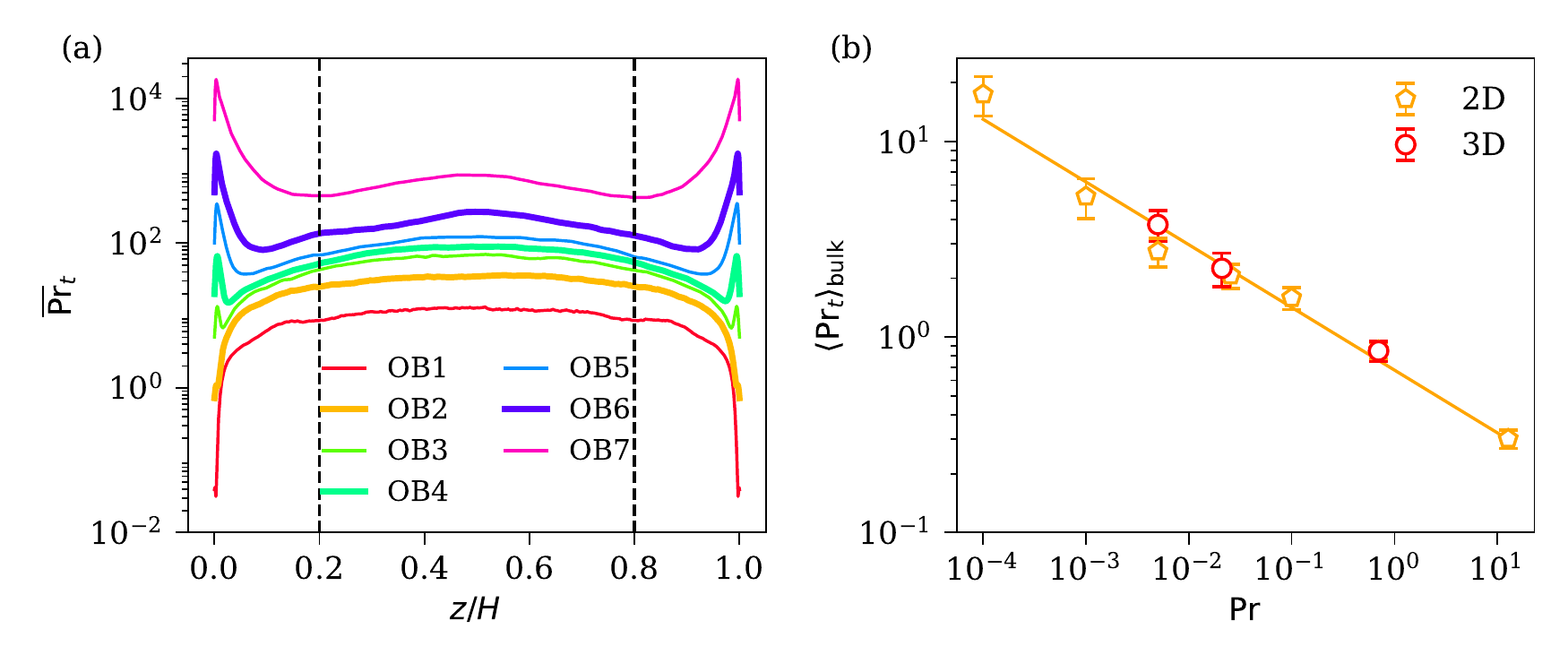}
\caption{(a) Vertical profiles of the turbulent Prandtl number in OB convection; $\overline{\mathrm{Pr}}_t(z)$ varies rapidly near the plates, whereas slowly in the bulk region, whose extent is indicated by dashed vertical lines. (b) The rescaled averaged $\mathrm{Pr}_t$ in the bulk region increases with decreasing $\mathrm{Pr}$. The error bars represent the standard deviation of $\overline{\mathrm{Pr}}_t$ in the bulk region.  Here $\langle \mathrm{Pr}_t \rangle_\mathrm{bulk}$ is obtained by using $c_\nu/c_\kappa = 0.0266$ in 2D, whereas $c_\nu/c_\kappa = 1.43$ in 3D such that it is approximately 0.85 for $\mathrm{Pr} = 0.7$. The solid line represents the best fit curve for the entire 2D data range, and is also a good fit for the 3D data.}
\label{fig:Prt_OB}
\end{figure}

The profiles of turbulent viscosity and turbulent thermal diffusivity are computed using profiles of Fig.~\ref{fig:profiles_comp_OB} as
\begin{eqnarray}
\overline\nu_t(z) & = & k^2_u(z)/\varepsilon_{u^{\prime}}(z), \label{eq:nut} \\
\overline\kappa_t(z) & = & k_u(z) k_T(z)/\varepsilon_{T^\prime}(z), \label{eq:kappat}
\end{eqnarray}
where we have used the overbar to denote the turbulent Prandtl number without  specifying  $c_\nu/c_\kappa$, whereas $\mathrm{Pr}_t$ without the overbar includes $c_\nu/c_\kappa$. The ratio $\overline{\mathrm{Pr}}_t(z) = \overline\nu_t(z)/\overline\kappa_t(z)$ as a function of $z$ is plotted in Fig.~\ref{fig:Prt_OB}(a). As mentioned earlier, we do not specify the coefficients $c_\nu$ and $c_\kappa$ and merely plot the ratio of the profiles shown in Fig.~\ref{fig:profiles_comp_OB}. 
Figure~\ref{fig:Prt_OB}(a) shows that $\overline{\mathrm{Pr}}_t(z)$ varies slowly in the bulk region. In the vicinity of the plates, $\overline{\mathrm{Pr}}_t$ increases with increasing distance from the plates, which is reasonable as the turbulent fluctuations strengthen near the plates, leading to higher turbulent transport of momentum and heat. The local maxima near the plates in Fig.~\ref{fig:Prt_OB}(a) are observed due to the peaks of $k_T(z)$ in Fig.~\ref{fig:profiles_comp_OB}(b). Note however that $\nu_t$ and $\kappa_t$ in RBC have been observed to scale as $z^3$ in the vicinity of the plates~\citep{Shishkina:PRL2015, Shishkina:PRF2017, Pandey:JFM2021, Tai:PRF2021}, which suggests the constancy of $\overline{\mathrm{Pr}}_t(z)$ in the near-wall region. As we are mainly concerned here with the behavior of $\overline{\mathrm{Pr}}_t$ in the bulk region, we do not explore further the near-wall variation of the turbulent Prandtl number.

To illustrate the variation of $\mathrm{Pr}_t$ with $\mathrm{Pr}$, we plot $\overline{\mathrm{Pr}}_t$ averaged in the bulk region as a function of $\mathrm{Pr}$ in Fig.~\ref{fig:Prt_OB}(b). We find that $\langle \mathrm{Pr}_t \rangle_\mathrm{bulk}$ increases with decreasing $\mathrm{Pr}$, which is qualitatively consistent with the observations in RBC in a cylindrical cell within the thermal BL region~\citep{Tai:PRF2021}. As we have not specified $c_\nu/c_\kappa$, only the variation of $\overline{\mathrm{Pr}}_t$ is of interest here. Here, $\langle \mathrm{Pr}_t \rangle_\mathrm{bulk}$ is obtained by using $c_\nu/c_\kappa = 0.0266$ such that it is approximately 0.85 for $\mathrm{Pr} = 0.7$. The value of $c_\nu/c_\kappa$ is 1.43 for the 3D cases. Similar data plotted in the figure from the 3D simulations also agree with the trend. Note that we did not include the 3D run at $\mathrm{Pr}=7$ which resulted in a Reynolds number of only 11, hence far from turbulent. In summary, all available OB data in the bulk obey the same scaling law. 

\begin{figure}
\begin{center}
\includegraphics[width=\textwidth]{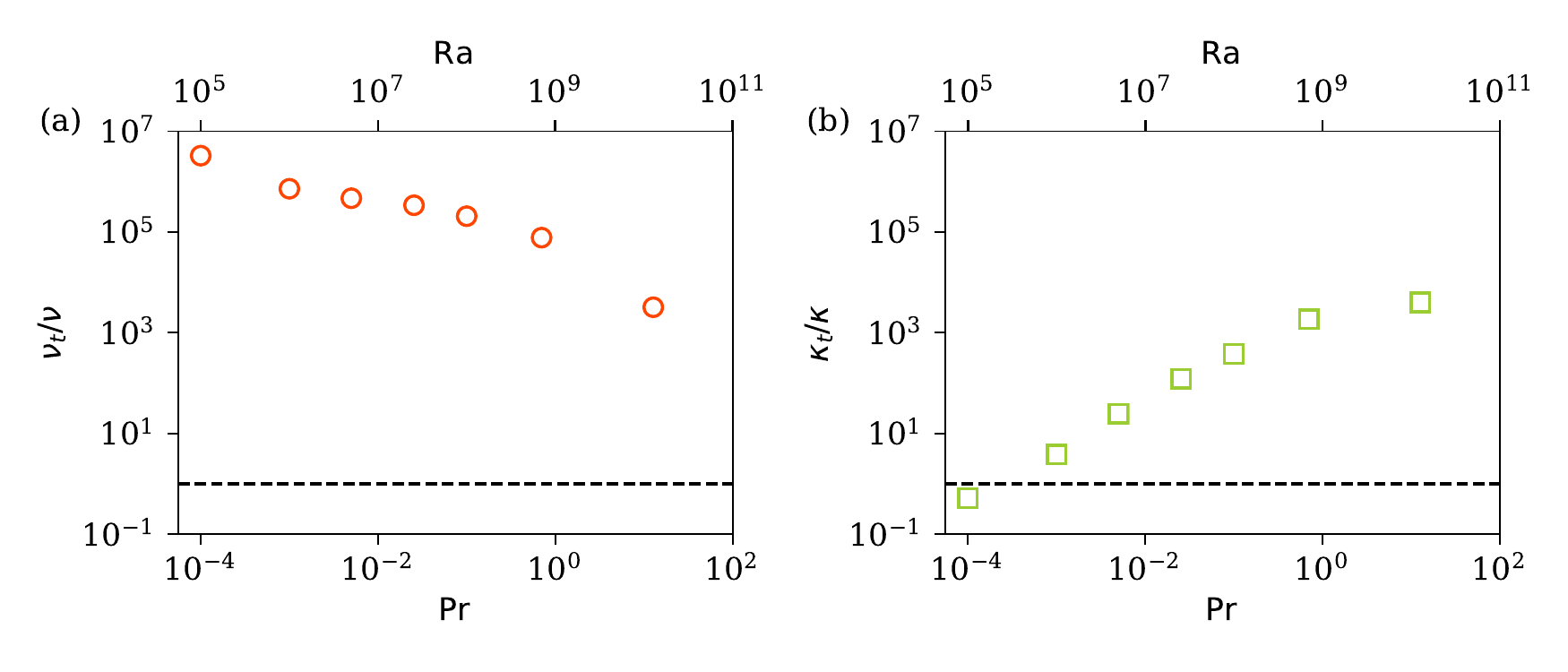}
\caption{Ratio of the turbulent and the molecular viscosity, $\nu_t/\nu$, in (a), and the turbulent and the molecular thermal diffusivity, $\kappa_t/\kappa$, in (b). Both ratios are shown as functions of $\mathrm{Pr}$ and $\mathrm{Ra}$ for the 2D OB simulations. Dashed horizontal lines help indicate the departure from unity. We observe $\nu_t \gg \nu$ for all cases, whereas the condition $\kappa_t \gg \kappa$ is satisfied for flows with $\mathrm{Pr} > 10^{-3}$.}
\label{fig:norm_turb_diffs}
\end{center}
\end{figure}

At this stage, it is also instructive to see how the magnitudes of the turbulent viscosity and turbulent thermal diffusivity individually differ from their molecular counterparts. Therefore, we compute the turbulent diffusivities using $c_\nu = 0.09, \, c_\kappa = 0.1$, and show in Fig.~\ref{fig:norm_turb_diffs} the ratios $\nu_t /\nu$ and $\kappa_t /\kappa$ for our OB simulations in 2D. Here, $\nu_t$ and $\kappa_t$ are bulk-averaged turbulent diffusivities. Figure~\ref{fig:norm_turb_diffs}(a) shows that the turbulent viscosity is much higher than the molecular viscosity for all cases explored, and is nearly three orders of magnitude larger than $\nu$ for $\mathrm{Pr} = 12.73$, with the contrast between $\nu_t$ and $\nu$ increasing further as $\mathrm{Pr}$ decreases; $\nu_t/\nu \approx 10^6$ for $\mathrm{Pr} = 10^{-4}$. On the other hand, Fig.~\ref{fig:norm_turb_diffs}(b) shows that the ratio $\kappa_t/\kappa$ decreases from nearly 5000 for $\mathrm{Pr} = 12.73$ to nearly 0.5 for $\mathrm{Pr} = 10^{-4}$. Thus, the turbulent and the molecular thermal diffusivities do not differ much when $\mathrm{Pr}\le 10^{-3}$ at the accessible Rayleigh numbers.

Our findings suggest that the disparity between the turbulent and molecular Prandtl numbers becomes very strong as $\mathrm{Pr}$ becomes low. If this trend continues for lower Prandtl numbers, the turbulent convective flow in the solar and stellar interiors would correspond to very high turbulent Prandtl numbers. Many nonlinear relations between $\mathrm{Pr}_t$ and $\mathrm{Pr}$ have been proposed in literature~\citep{Reynolds:IJHMT1975, Abe:IJHMT2019}. The data in Fig.~\ref{fig:Prt_OB}(b) suggest that $\langle \mathrm{Pr}_t \rangle_\mathrm{bulk}$ can be fitted as a power law, and the best fit yields $\langle \mathrm{Pr}_t \rangle_\mathrm{bulk} = (0.68 \pm 0.03)\mathrm{Pr}^{-0.32 \pm 0.02}$. \citet{Abe:IJHMT2019} performed channel flow simulations for $\mathrm{Pr} = 0.7$ and 0.025 to compute $\mathrm{Pr}_t$ using the flux-gradient methods, and also observed an increasing $\mathrm{Pr}_t$ with decreasing $\mathrm{Pr}$. This result is consistent with less detailed results of \citet{Reynolds:IJHMT1975, Jischa:IJHMT1979, Bricteux:NSE2012, Tai:PRF2021}.
 
What does a decreasing $\mathrm{Pr}_t$ with increasing $\mathrm{Pr}$ imply? Note that $\mathrm{Pr}_t$ is a ratio of the turbulent momentum flux compared to the turbulent heat flux for unit mean velocity and temperature gradients. Thus, an increasing turbulent Prandtl number indicates that turbulent fluctuations transport momentum more effectively than heat. This is not counter-intuitive because, with decreasing $\mathrm{Pr}$, the momentum transport ($\mathrm{Re}$) increases whereas the heat transport ($\mathrm{Nu}$) decreases in RBC.

\section{Turbulent Prandtl number in non-Oberbeck-Boussinesq convection}
\label{sec:NOB}

\subsection{Flow structure and global transport of heat and momentum}
\label{sec:str_NOB}

\begin{figure}
\includegraphics[width=0.9\textwidth]{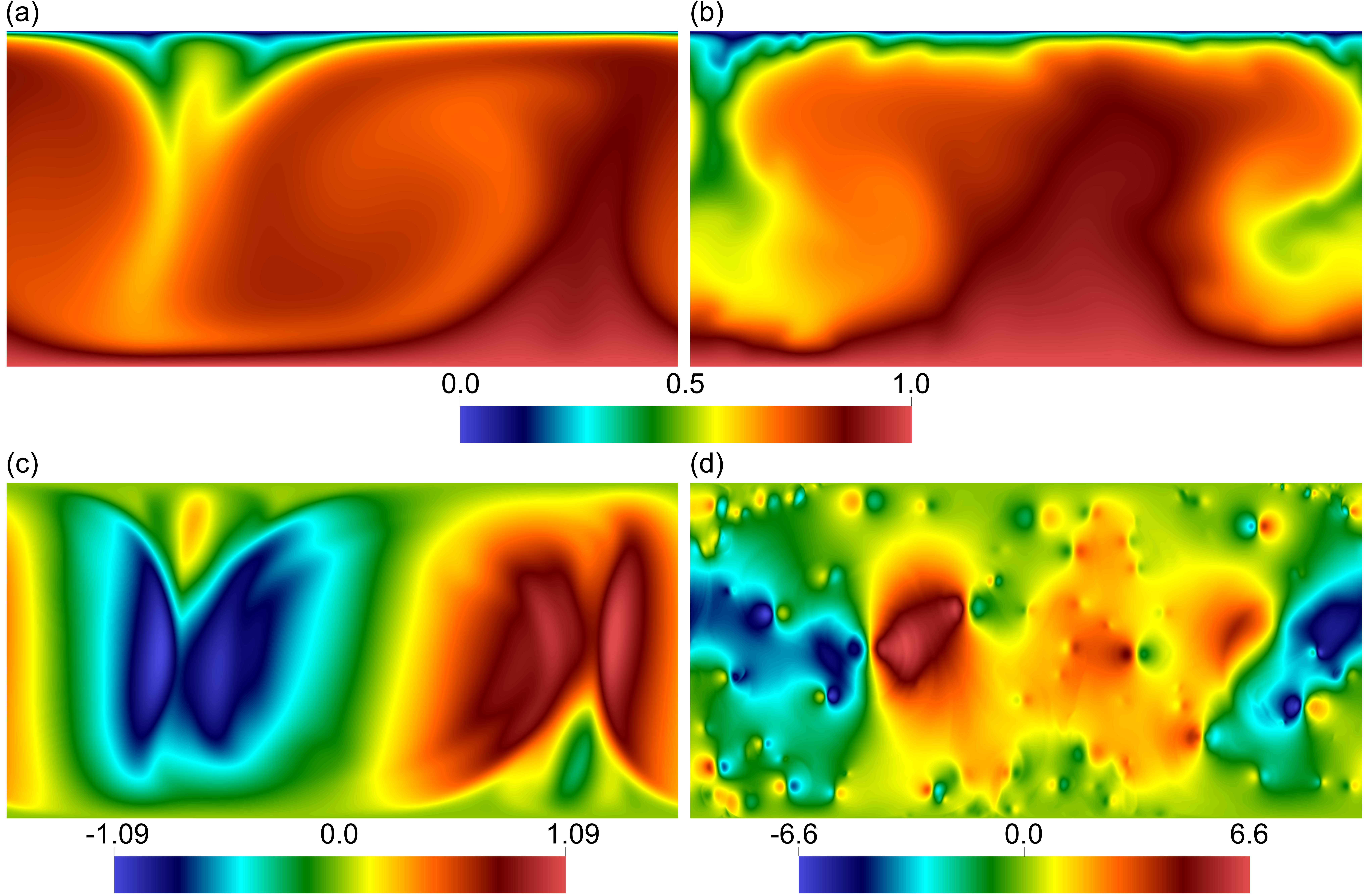}
\caption{Instantaneous temperature (a, b) and vertical velocity (c, d) fields for NOB2 (a, c) and NOB6 (b, d) simulations. Characteristic thickness of the thermal structures are larger in (b) than in (a). The velocity field in (c) is smoother, whereas more intermittent structures can be observed in (d) due to a much higher Reynolds number in the latter (see Table~\ref{table:details_NOB}).}
\label{fig:T_uz_NOB}
\end{figure}

We plot the instantaneous temperature and vertical velocity fields for NOB2 and NOB6 in Fig.~\ref{fig:T_uz_NOB}. The figure reveals that the hot plumes emanating from the bottom plate are coarser compared to their colder counterparts from the top plate. This is due to the difference between the thermal diffusivities at the top and bottom plates, mentioned earlier. 
We also observe from Fig.~\ref{fig:T_uz_NOB} that the average temperature is larger than 0.5, which is due to the specified positive correlation between $\kappa$ and $T$~\citep{Sameen:EPL2009}. The finding is in agreement with the observations of Fig.~\ref{fig:T_Pr_z}(a). Due to a larger diffusivity of the hotter plumes, their thermal diffusion time scale ($= H^2/\kappa(T)$) is shorter. As a result, hotter structures diffuse quickly compared to the colder ones and occupy a larger fraction of the flow, thus resulting in a higher mean temperature. 

\begin{figure}
\includegraphics[width=\textwidth]{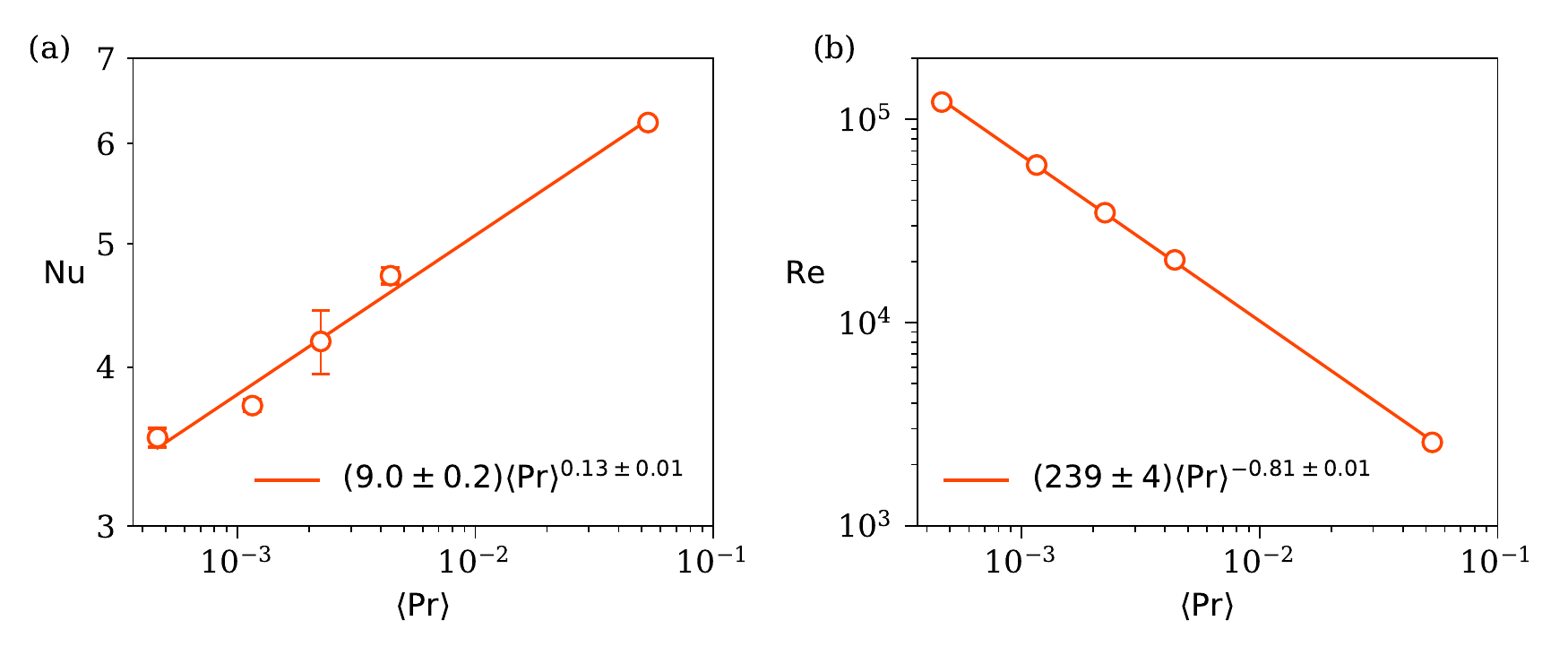}
\caption{(a) The Nusselt number and (b) the Reynolds number as a function of the globally-averaged Prandtl number for NOB simulations with $\kappa_2(T)$. The powerlaw exponents in both the $\mathrm{Nu}$ and $\mathrm{Re}$ scalings are in the range of exponents observed in OB convection.}
\label{fig:Nu_Re_NOB}
\end{figure}

To see how the global heat transport varies with $\mathrm{Pr}$ in NOB simulations, we compute the Nusselt number as $\mathrm{Nu} = (\langle u_z T \rangle_{A,t} + J_d)/J_c$, where $J_d = -\langle \kappa(T) \partial T/\partial z \rangle_{A,t}$ is the heat flux due to molecular diffusion and $J_c$ is the heat flux in the conduction state. We refer to \citet{Pandey:APJ2021} for a detailed discussion on the computation of heat and momentum transport, as well as dissipation rates in this NOB flow. The values of $\mathrm{Nu}$ are listed in Table~\ref{table:details_NOB}. The variation of $\mathrm{Nu}$ with the mean Prandtl number $\langle \mathrm{Pr} \rangle$ for simulations with $\kappa_2(T)$, shown in Fig.~\ref{fig:Nu_Re_NOB}(a), reveals that $\mathrm{Nu}$ increases as $\langle \mathrm{Pr} \rangle^{0.13 \pm 0.01}$. The Nusselt numbers for all simulations are of the order of unity, thus indicating that the molecular diffusion is significant in transporting heat, even when the flow has a rich turbulent structure. The power law exponent of $\mathrm{Nu-Pr}$ scaling is in the range observed in OB convection~\citep{Verzicco:JFM1999, Grossmann:JFM2000, Scheel:PRF2017, Pandey:EPL2021}.

As for OB convection, we also compute the Nusselt number using the viscous and thermal dissipation rates as $\mathrm{Nu}_{\varepsilon_u} = [(\gamma g)^{-1} \langle \varepsilon_u \rangle_{A,t} + J_d]/J_c$ and $\mathrm{Nu}_{\varepsilon_T} = \langle \varepsilon_T \rangle_{A,t} H/J_c \Delta T$~\citep{Pandey:APJ2021}, and list them in Table~\ref{table:details_NOB}. We find that the Nusselt number computed using the three methods agree reasonably well, further indicating that the NOB flows are sufficiently resolved. We estimate the momentum transport in the NOB simulations by computing the Reynolds number as $\mathrm{Re} = u_\mathrm{rms} \sqrt{\mathrm{Gr}}$, and plot it as a function of the mean Prandtl number in Fig.~\ref{fig:Nu_Re_NOB}(b). We find that $\mathrm{Re}$ decreases as $\langle \mathrm{Pr} \rangle^{-0.81 \pm 0.01}$. The $\mathrm{Pr}$-dependence of $\mathrm{Re}$ in the NOB flows is qualitatively similar to that in OB convection~\citep{Verzicco:JFM1999, Yang:JFM2021, Li:JFM2021}. 

\subsection{The turbulent Prandtl number in the non-Oberbeck-Boussinesq case}
\label{sec:Prt_NOB}

\begin{figure}
\includegraphics[width=0.9\textwidth]{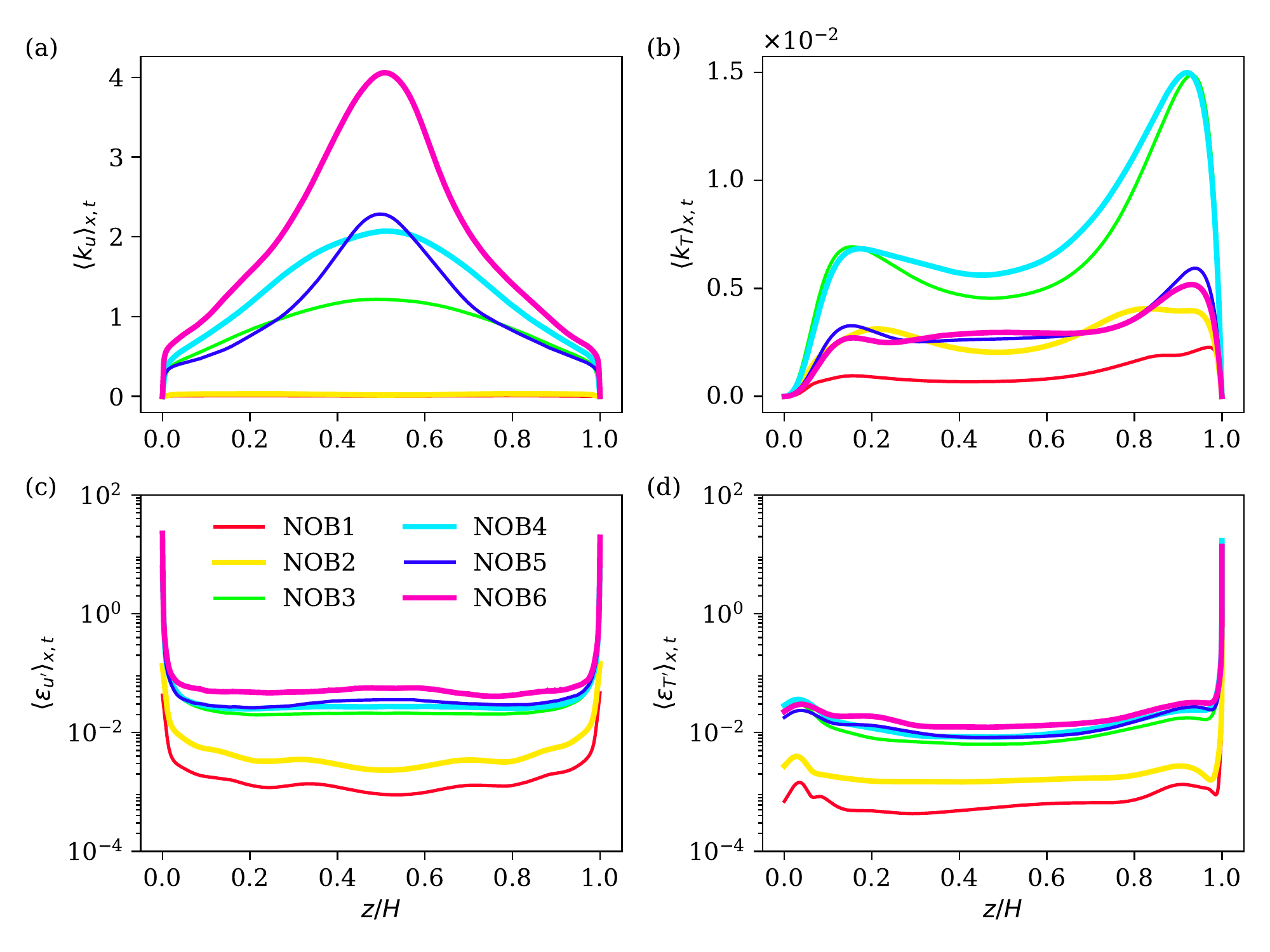}
\caption{Vertical profiles of the (a) turbulent kinetic energy, (b) thermal variance, (c) turbulent viscous dissipation rate, and (d) turbulent thermal dissipation rate for NOB simulations. In contrast to OB simulations, the profiles in (b) and (d) are asymmetric about $z = H/2$.}
\label{fig:profiles_comp_NOB}
\end{figure}

We show again vertical profiles of $k_u, k_T, \varepsilon_{u^\prime}, \varepsilon_{T^\prime}$ in Fig.~\ref{fig:profiles_comp_NOB}. The profiles are fairly symmetric with respect to the midplane $z = H/2$, consistent with~\citet{Pandey:APJ2021}: quantities related to the velocity field are not affected much by using only a temperature-dependent thermal diffusivity.
However, Fig.~\ref{fig:profiles_comp_NOB} (b,d) show that $k_T$ and $\varepsilon_{T^\prime}$ are asymmetric with respect to midplane, and the thermal dissipation rate increases rapidly towards the top plate. This is because of a larger $\partial T/\partial z$ in the region near the top plate. An asymmetric $k_T(z)$ is consistent with unequal thicknesses of the thermal boundary layers at the two plates~\citep{Pandey:APJ2021}. 

\begin{figure}
\includegraphics[width=\textwidth]{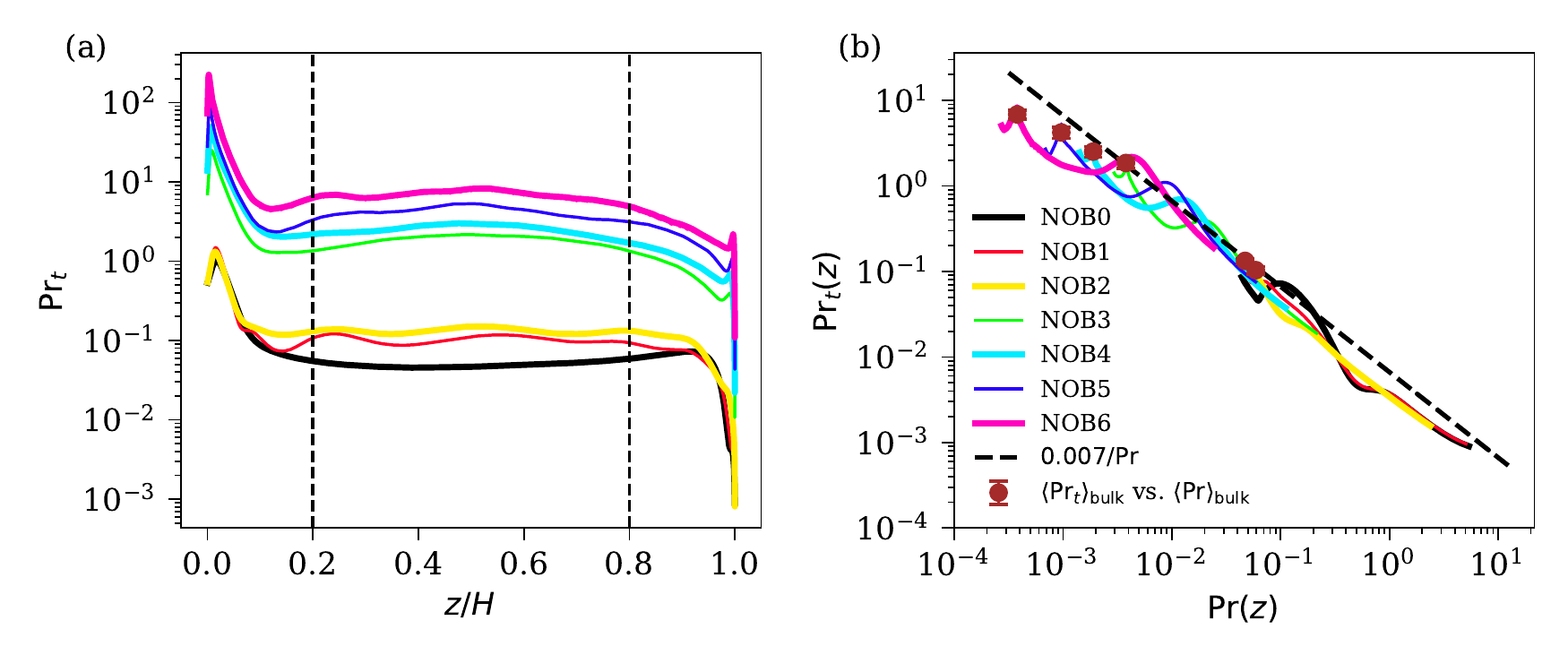}
\caption{ (a) Turbulent Prandtl number as a function of $z$ and (b) $\mathrm{Pr}_t(z)$ as a function of $\mathrm{Pr} (z)$ for $z/H \in [0.10  \,\, 0.9999]$ for the NOB simulations. NOB0 corresponds to the results from our simulation in a 3D rectangular box of $\Gamma = 16$ performed for $\kappa_1(T)$ (taken from~\citet{Pandey:APJ2021}). Panel (b) indicates that $\mathrm{Pr}_t$ nearly scales as $1/\mathrm{Pr}$. Further, the variation of $\mathrm{Pr}_t$ with $\mathrm{Pr}$ is qualitatively similar for both the NOB0 and NOB1 flows. Markers indicate the turbulent Prandtl number averaged in the bulk region between $z = 0.2H$ and $z = 0.8H$ as a function of the averaged-$\mathrm{Pr}$ in the same bulk region.}
\label{fig:Prt_z_Pr_NOB}
\end{figure}

Using the profiles shown in Fig.~\ref{fig:profiles_comp_NOB}, we compute $\mathrm{Pr}_t$ in each horizontal plane as $\mathrm{Pr}_t(z) = \nu_t(z)/\kappa_t(z)$ and show them in Fig.~\ref{fig:Prt_z_Pr_NOB}(a). For comparison, we also show the results from our simulation for the 3D box of $\Gamma = 16$ (NOB0, black curves in both panels, from~\citet{Pandey:APJ2021}), for the same diffusivity profile and the same governing parameters as for NOB1. Note that $\mathrm{Pr}_t$ in Fig.~\ref{fig:Prt_z_Pr_NOB} is already rescaled using the same $c_\nu/c_\kappa = 0.0266$ as used for the OB cases in Sec.~\ref{sec:OB}. Note also that, in contrast to the OB case, the profiles of $\mathrm{Pr}_t$ are asymmetric with respect to the midplane. The turbulent Prandtl number varies rapidly near the plates, with the lowest $\mathrm{Pr}_t(z)$ at the top. 

Figure~\ref{fig:Prt_z_Pr_NOB}(a) shows that the variation of $\mathrm{Pr}_t$ near the plates is very similar for NOB0 and NOB1. In the bulk region, however, the magnitude of $\mathrm{Pr}_t$ in 3D flow is consistently lower than for the 2D flow. This is plausible because the turbulent momentum transport in a 2D convective flow is higher than in the 3D flow with the same control parameters, whereas the heat transport in the two flows does not differ as much~\citep{Pandey:JFM2021}. For instance, we observe $\mathrm{Nu} = 5.4$ and $\mathrm{Re} = 1100$ for NOB0~\citep{Pandey:APJ2021} compared to $6.6$ and $2048$ for NOB1 (see Table~\ref{table:details_NOB}). Since $\mathrm{Pr}_t$ is the ratio of the momentum and heat transports due to turbulent fluctuations for unit mean velocity and temperature gradients, a relatively higher momentum transport in NOB1 leads to a higher $\mathrm{Pr}_t$ in NOB1 than in NOB0.

We observe from Fig.~\ref{fig:T_Pr_z}(b) that the major change in the molecular Prandtl number occurs in the thin thermal BL region near the top plate. Therefore, to see the variation of $\mathrm{Pr}_t$ with $\mathrm{Pr}$ in our NOB flows, we plot $\mathrm{Pr}_t(z)$ as a function of $\mathrm{Pr} (z)$ for $z/H \in [0.10 \, \, 0.9999]$ in Fig.~\ref{fig:Prt_z_Pr_NOB}(b), which shows that $\mathrm{Pr}_t$ varies nearly inversely with $\mathrm{Pr}$, this being steeper than $\mathrm{Pr}_t \sim \mathrm{Pr}^{-0.31}$ observed for OB simulations. 
\section{Final Discussion} \label{sec:conclu}

\subsection{Summary}
We have studied the variation of the turbulent Prandtl number with respect to the molecular Prandtl number using direct numerical simulations of thermal convection in OB and NOB settings. For the 2D case, we have $\Gamma = 2$, $\mathrm{Pr}$ between 12.73 and $10^{-4}$ and $\mathrm{Gr} = 10^9$. We also computed large aspect ratio 3D simulations. We computed $\mathrm{Pr}_t$ within the $k-\varepsilon$ framework. We found that the averaged turbulent Prandtl number in the bulk region $\langle \mathrm{Pr}_t \rangle_\mathrm{bulk} \sim \mathrm{Pr}^{-0.31}$, over five orders of magnitude of the molecular Prandtl number, $\mathrm{Pr}$. This result suggests that low-Prandtl-number turbulent convection behave effectively as high-Prandtl-number flows.

Since OB approximations do not apply to natural convective flows~\citep{Schumacher:RMP2020}, we explored NOB convection by varying the thermal diffusivity with temperature, which breaks the top-down symmetry of the flow. We performed DNS for a fixed $\mathrm{Ra}$ at the top plate and varied $\mathrm{Pr}$ at the top plate, with the mean $\mathrm{Pr}$ over the entire domain reaching up to $5 \times 10^{-4}$. This simple NOB flow exhibits several properties of fully compressible flows~\citep{Pandey:APJ2021}, and the estimated $\langle \mathrm{Pr}_t \rangle_\mathrm{bulk}$ has the behavior $\langle \mathrm{Pr}_t \rangle_\mathrm{bulk} \sim \mathrm{Pr}^{-1}$, a much stronger dependence than in OB flows. 

\subsection{Outlook and closure}
Our findings indicate that the convection processes in solar and stellar interiors, which correspond to extremely low-$\mathrm{Pr}$ flows, may be characterized by very high turbulent Prandtl numbers. Indeed, our results have important implications for the modeling of low-Prandtl-number convection anywhere.  

For one set of conditions with varying thermal diffusivity, we found the result that the turbulent Prandtl number varies inversely with the molecular Prandtl number. (We have also explored the temperature dependence of viscosity, whose preliminary assessment is essentially similar.) The smallest molecular Prandtl number in these simulations is quite small, in fact the smallest ever, but it is still not as small as in the Sun; but, given the simplicity of the fit, perhaps we can extrapolate the trend and make some tentative deductions. One such deduction is that the turbulent Prandtl number in the Sun’s convective region is of the order of $\mathrm{Pr}_t\sim 10^4$. This large value has important implications. 

Now consider the case where the initial velocity and temperature fields are represented by two cosine waves. And the two waves are partially correlated. A short time later, since the effective viscosity is 10,000 times larger than the effective thermal diffusivity, the velocity fluctuations get smoothed out whereas the temperature fluctuations remain unmixed. We thus have the case of temperature structures being advected essentially by a structureless, random velocity field. This velocity field is much smaller in magnitude than one might expect from standard phenomenology. In our opinion, this offers an explanation for how large amounts of thermal energy can be transported despite small velocities, a conundrum articulated in \citet{Hanasoge:PNAS2012}. This also opens up our inquiry towards new theoretical ideas, e.g., of temperature plumes persisting through the entire convection zone despite the very high Rayleigh numbers in the Sun. 

As a final caveat, we mention that rotation is an important factor of structure formation and turbulent transport that needs to be taken into account in order to get a more realistic picture of convection in the Sun. Rotation, however, mostly affects the meso- and larger scales which would include supergranules and hypothetical giant cells; see simulations of solar convection in Refs. \citep{Featherstone:APJL2016, Karak:POF2018, Vasil:PNAS2021}. As most of our study in this paper is concerned with small-scale turbulence properties, which would correspond to scales below the solar granules diameter with $\sim 10^3$ km, we focused in the present work to two-dimensional simulations and do not include the effects of rotation. This work is proceeding at this time.


\acknowledgements{This research was carried out on the High Performance Computing resources at New York University Abu Dhabi.}

%


\begin{thebibliography}{70}%
\makeatletter
\providecommand \@ifxundefined [1]{%
 \@ifx{#1\undefined}
}%
\providecommand \@ifnum [1]{%
 \ifnum #1\expandafter \@firstoftwo
 \else \expandafter \@secondoftwo
 \fi
}%
\providecommand \@ifx [1]{%
 \ifx #1\expandafter \@firstoftwo
 \else \expandafter \@secondoftwo
 \fi
}%
\providecommand \natexlab [1]{#1}%
\providecommand \enquote  [1]{``#1''}%
\providecommand \bibnamefont  [1]{#1}%
\providecommand \bibfnamefont [1]{#1}%
\providecommand \citenamefont [1]{#1}%
\providecommand \href@noop [0]{\@secondoftwo}%
\providecommand \href [0]{\begingroup \@sanitize@url \@href}%
\providecommand \@href[1]{\@@startlink{#1}\@@href}%
\providecommand \@@href[1]{\endgroup#1\@@endlink}%
\providecommand \@sanitize@url [0]{\catcode `\\12\catcode `\$12\catcode
  `\&12\catcode `\#12\catcode `\^12\catcode `\_12\catcode `\%12\relax}%
\providecommand \@@startlink[1]{}%
\providecommand \@@endlink[0]{}%
\providecommand \url  [0]{\begingroup\@sanitize@url \@url }%
\providecommand \@url [1]{\endgroup\@href {#1}{\urlprefix }}%
\providecommand \urlprefix  [0]{URL }%
\providecommand \Eprint [0]{\href }%
\providecommand \doibase [0]{https://doi.org/}%
\providecommand \selectlanguage [0]{\@gobble}%
\providecommand \bibinfo  [0]{\@secondoftwo}%
\providecommand \bibfield  [0]{\@secondoftwo}%
\providecommand \translation [1]{[#1]}%
\providecommand \BibitemOpen [0]{}%
\providecommand \bibitemStop [0]{}%
\providecommand \bibitemNoStop [0]{.\EOS\space}%
\providecommand \EOS [0]{\spacefactor3000\relax}%
\providecommand \BibitemShut  [1]{\csname bibitem#1\endcsname}%
\let\auto@bib@innerbib\@empty
\bibitem [{\citenamefont {Schumacher}\ and\ \citenamefont
  {Sreenivasan}(2020)}]{Schumacher:RMP2020}%
  \BibitemOpen
  \bibfield  {author} {\bibinfo {author} {\bibfnamefont {J.}~\bibnamefont
  {Schumacher}}\ and\ \bibinfo {author} {\bibfnamefont {K.~R.}\ \bibnamefont
  {Sreenivasan}},\ }\bibfield  {title} {\bibinfo {title} {Colloquium: Unusual
  dynamics of convection in the sun},\ }\href
  {https://doi.org/10.1103/RevModPhys.92.041001} {\bibfield  {journal}
  {\bibinfo  {journal} {Rev. Mod. Phys.}\ }\textbf {\bibinfo {volume} {92}},\
  \bibinfo {pages} {041001} (\bibinfo {year} {2020})}\BibitemShut {NoStop}%
\bibitem [{\citenamefont {Sreenivasan}(2019)}]{Sreenivasan:PNAS2019}%
  \BibitemOpen
  \bibfield  {author} {\bibinfo {author} {\bibfnamefont {K.~R.}\ \bibnamefont
  {Sreenivasan}},\ }\bibfield  {title} {\bibinfo {title} {Turbulent mixing: A
  perspective},\ }\href {https://doi.org/10.1073/pnas.1800463115} {\bibfield
  {journal} {\bibinfo  {journal} {Proc. Natl. Acad. Sci. USA}\ }\textbf
  {\bibinfo {volume} {116}},\ \bibinfo {pages} {18175} (\bibinfo {year}
  {2019})},\ \Eprint
  {https://arxiv.org/abs/https://www.pnas.org/content/116/37/18175.full.pdf}
  {https://www.pnas.org/content/116/37/18175.full.pdf} \BibitemShut {NoStop}%
\bibitem [{\citenamefont {{Hanasoge}}\ \emph {et~al.}(2016)\citenamefont
  {{Hanasoge}}, \citenamefont {{Gizon}},\ and\ \citenamefont
  {{Sreenivasan}}}]{Hanasoge:ARFM2016}%
  \BibitemOpen
  \bibfield  {author} {\bibinfo {author} {\bibfnamefont {S.}~\bibnamefont
  {{Hanasoge}}}, \bibinfo {author} {\bibfnamefont {L.}~\bibnamefont
  {{Gizon}}},\ and\ \bibinfo {author} {\bibfnamefont {K.~R.}\ \bibnamefont
  {{Sreenivasan}}},\ }\bibfield  {title} {\bibinfo {title} {Seismic sounding of
  convection in the sun},\ }\href
  {https://doi.org/10.1146/annurev-fluid-122414-034534} {\bibfield  {journal}
  {\bibinfo  {journal} {Annu. Rev. Fluid Mech.}\ }\textbf {\bibinfo {volume}
  {48}},\ \bibinfo {pages} {191} (\bibinfo {year} {2016})}\BibitemShut
  {NoStop}%
\bibitem [{\citenamefont {Sreenivasan}(1998)}]{Sreenivasan:1998}%
  \BibitemOpen
  \bibfield  {author} {\bibinfo {author} {\bibfnamefont {K.~R.}\ \bibnamefont
  {Sreenivasan}},\ }\bibinfo {title} {Helium flows at ultra-high {R}eynolds and
  {R}ayleigh numbers: Opportunities and challenges},\ in\ \href
  {https://doi.org/10.1007/978-1-4612-2230-9_2} {\emph {\bibinfo {booktitle}
  {Flow at Ultra-High {R}eynolds and {R}ayleigh Numbers: A Status Report}}},\
  \bibinfo {editor} {edited by\ \bibinfo {editor} {\bibfnamefont {R.~J.}\
  \bibnamefont {Donnelly}}\ and\ \bibinfo {editor} {\bibfnamefont {K.~R.}\
  \bibnamefont {Sreenivasan}}}\ (\bibinfo  {publisher} {Springer New York},\
  \bibinfo {address} {New York, NY},\ \bibinfo {year} {1998})\ pp.\ \bibinfo
  {pages} {29--51}\BibitemShut {NoStop}%
\bibitem [{\citenamefont {Ahlers}\ \emph {et~al.}(2009)\citenamefont {Ahlers},
  \citenamefont {Grossmann},\ and\ \citenamefont {Lohse}}]{Ahlers:RMP2009}%
  \BibitemOpen
  \bibfield  {author} {\bibinfo {author} {\bibfnamefont {G.}~\bibnamefont
  {Ahlers}}, \bibinfo {author} {\bibfnamefont {S.}~\bibnamefont {Grossmann}},\
  and\ \bibinfo {author} {\bibfnamefont {D.}~\bibnamefont {Lohse}},\ }\bibfield
   {title} {\bibinfo {title} {Heat transfer and large scale dynamics in
  turbulent {R}ayleigh-{B}\'enard convection},\ }\href
  {https://doi.org/10.1103/RevModPhys.81.503} {\bibfield  {journal} {\bibinfo
  {journal} {Rev. Mod. Phys.}\ }\textbf {\bibinfo {volume} {81}},\ \bibinfo
  {pages} {503} (\bibinfo {year} {2009})}\BibitemShut {NoStop}%
\bibitem [{\citenamefont {{Chill\`{a}}}\ and\ \citenamefont
  {{Schumacher}}(2012)}]{Chilla:EPJE2012}%
  \BibitemOpen
  \bibfield  {author} {\bibinfo {author} {\bibfnamefont {F.}~\bibnamefont
  {{Chill\`{a}}}}\ and\ \bibinfo {author} {\bibfnamefont {J.}~\bibnamefont
  {{Schumacher}}},\ }\bibfield  {title} {\bibinfo {title} {New perspectives in
  turbulent {R}ayleigh-{B}{\'e}nard convection},\ }\href
  {https://doi.org/10.1140/epje/i2012-12058-1} {\bibfield  {journal} {\bibinfo
  {journal} {Eur. Phys. J. E}\ }\textbf {\bibinfo {volume} {35}},\ \bibinfo
  {pages} {58} (\bibinfo {year} {2012})}\BibitemShut {NoStop}%
\bibitem [{\citenamefont {Verma}\ \emph {et~al.}(2017)\citenamefont {Verma},
  \citenamefont {Kumar},\ and\ \citenamefont {Pandey}}]{Verma:NJP2017}%
  \BibitemOpen
  \bibfield  {author} {\bibinfo {author} {\bibfnamefont {M.~K.}\ \bibnamefont
  {Verma}}, \bibinfo {author} {\bibfnamefont {A.}~\bibnamefont {Kumar}},\ and\
  \bibinfo {author} {\bibfnamefont {A.}~\bibnamefont {Pandey}},\ }\bibfield
  {title} {\bibinfo {title} {Phenomenology of buoyancy-driven turbulence:
  recent results},\ }\href {http://stacks.iop.org/1367-2630/19/i=2/a=025012}
  {\bibfield  {journal} {\bibinfo  {journal} {New J. Phys.}\ }\textbf {\bibinfo
  {volume} {19}},\ \bibinfo {pages} {025012} (\bibinfo {year}
  {2017})}\BibitemShut {NoStop}%
\bibitem [{\citenamefont {Verma}(2018)}]{Verma:book2018}%
  \BibitemOpen
  \bibfield  {author} {\bibinfo {author} {\bibfnamefont {M.~K.}\ \bibnamefont
  {Verma}},\ }\href {https://doi.org/10.1142/10928} {\emph {\bibinfo {title}
  {Physics of Buoyant Flows}}}\ (\bibinfo  {publisher} {World Scientific},\
  \bibinfo {address} {Sigapore},\ \bibinfo {year} {2018})\ \Eprint
  {https://arxiv.org/abs/https://www.worldscientific.com/doi/pdf/10.1142/10928}
  {https://www.worldscientific.com/doi/pdf/10.1142/10928} \BibitemShut
  {NoStop}%
\bibitem [{\citenamefont {{Schubert}}\ \emph {et~al.}(2001)\citenamefont
  {{Schubert}}, \citenamefont {{Turcotte}},\ and\ \citenamefont
  {{Olson}}}]{Schubert:book2001}%
  \BibitemOpen
  \bibfield  {author} {\bibinfo {author} {\bibfnamefont {G.}~\bibnamefont
  {{Schubert}}}, \bibinfo {author} {\bibfnamefont {D.~L.}\ \bibnamefont
  {{Turcotte}}},\ and\ \bibinfo {author} {\bibfnamefont {P.}~\bibnamefont
  {{Olson}}},\ }\href@noop {} {\emph {\bibinfo {title} {Mantle Convection in
  the Earth and Planets}}}\ (\bibinfo  {publisher} {Cambridge University
  Press},\ \bibinfo {address} {Cambridge, UK},\ \bibinfo {year}
  {2001})\BibitemShut {NoStop}%
\bibitem [{\citenamefont {Yakhot}\ \emph {et~al.}(1987)\citenamefont {Yakhot},
  \citenamefont {Orszag},\ and\ \citenamefont {Yakhot}}]{Yakhot:IJHMT1987}%
  \BibitemOpen
  \bibfield  {author} {\bibinfo {author} {\bibfnamefont {V.}~\bibnamefont
  {Yakhot}}, \bibinfo {author} {\bibfnamefont {S.~A.}\ \bibnamefont {Orszag}},\
  and\ \bibinfo {author} {\bibfnamefont {A.}~\bibnamefont {Yakhot}},\
  }\bibfield  {title} {\bibinfo {title} {Heat transfer in turbulent fluids -
  {I}. pipe flow},\ }\href
  {https://doi.org/https://doi.org/10.1016/0017-9310(87)90057-3} {\bibfield
  {journal} {\bibinfo  {journal} {Int. J. Heat Mass Transfer}\ }\textbf
  {\bibinfo {volume} {30}},\ \bibinfo {pages} {15} (\bibinfo {year}
  {1987})}\BibitemShut {NoStop}%
\bibitem [{\citenamefont {Oti\'c}\ and\ \citenamefont
  {Gr\"otzbach}(2007)}]{Otic:NSE2007}%
  \BibitemOpen
  \bibfield  {author} {\bibinfo {author} {\bibfnamefont {I.}~\bibnamefont
  {Oti\'c}}\ and\ \bibinfo {author} {\bibfnamefont {G.}~\bibnamefont
  {Gr\"otzbach}},\ }\bibfield  {title} {\bibinfo {title} {Turbulent heat flux
  and temperature variance dissipation rate in natural convection in
  lead-bismuth},\ }\href {https://doi.org/10.13182/NSE07-A2679} {\bibfield
  {journal} {\bibinfo  {journal} {Nucl. Sci. Eng.}\ }\textbf {\bibinfo {volume}
  {155}},\ \bibinfo {pages} {489} (\bibinfo {year} {2007})},\ \Eprint
  {https://arxiv.org/abs/https://doi.org/10.13182/NSE07-A2679}
  {https://doi.org/10.13182/NSE07-A2679} \BibitemShut {NoStop}%
\bibitem [{\citenamefont {Li}(2019)}]{Li:AR2019}%
  \BibitemOpen
  \bibfield  {author} {\bibinfo {author} {\bibfnamefont {D.}~\bibnamefont
  {Li}},\ }\bibfield  {title} {\bibinfo {title} {Turbulent {P}randtl number in
  the atmospheric boundary layer - where are we now?},\ }\href
  {https://doi.org/https://doi.org/10.1016/j.atmosres.2018.09.015} {\bibfield
  {journal} {\bibinfo  {journal} {Atmos. Res.}\ }\textbf {\bibinfo {volume}
  {216}},\ \bibinfo {pages} {86 } (\bibinfo {year} {2019})}\BibitemShut
  {NoStop}%
\bibitem [{\citenamefont {Bricteux}\ \emph {et~al.}(2012)\citenamefont
  {Bricteux}, \citenamefont {Duponcheel}, \citenamefont {Winckelmans},
  \citenamefont {Tiselj},\ and\ \citenamefont
  {Bartosiewicz}}]{Bricteux:NSE2012}%
  \BibitemOpen
  \bibfield  {author} {\bibinfo {author} {\bibfnamefont {L.}~\bibnamefont
  {Bricteux}}, \bibinfo {author} {\bibfnamefont {M.}~\bibnamefont
  {Duponcheel}}, \bibinfo {author} {\bibfnamefont {G.}~\bibnamefont
  {Winckelmans}}, \bibinfo {author} {\bibfnamefont {I.}~\bibnamefont
  {Tiselj}},\ and\ \bibinfo {author} {\bibfnamefont {Y.}~\bibnamefont
  {Bartosiewicz}},\ }\bibfield  {title} {\bibinfo {title} {Direct and large
  eddy simulation of turbulent heat transfer at very low {P}randtl number:
  Application to lead–bismuth flows},\ }\href
  {https://doi.org/https://doi.org/10.1016/j.nucengdes.2011.07.010} {\bibfield
  {journal} {\bibinfo  {journal} {Nucl. Eng. Des.}\ }\textbf {\bibinfo {volume}
  {246}},\ \bibinfo {pages} {91} (\bibinfo {year} {2012})}\BibitemShut
  {NoStop}%
\bibitem [{\citenamefont {Abe}\ and\ \citenamefont
  {Antonia}(2019)}]{Abe:IJHMT2019}%
  \BibitemOpen
  \bibfield  {author} {\bibinfo {author} {\bibfnamefont {H.}~\bibnamefont
  {Abe}}\ and\ \bibinfo {author} {\bibfnamefont {R.~A.}\ \bibnamefont
  {Antonia}},\ }\bibfield  {title} {\bibinfo {title} {Mean temperature
  calculations in a turbulent channel flow for air and mercury},\ }\href
  {https://doi.org/https://doi.org/10.1016/j.ijheatmasstransfer.2018.11.100}
  {\bibfield  {journal} {\bibinfo  {journal} {Int. J. Heat Mass Transfer}\
  }\textbf {\bibinfo {volume} {132}},\ \bibinfo {pages} {1152} (\bibinfo {year}
  {2019})}\BibitemShut {NoStop}%
\bibitem [{\citenamefont {{R}eynolds}(1975)}]{Reynolds:IJHMT1975}%
  \BibitemOpen
  \bibfield  {author} {\bibinfo {author} {\bibfnamefont {A.}~\bibnamefont
  {{R}eynolds}},\ }\bibfield  {title} {\bibinfo {title} {The prediction of
  turbulent {P}randtl and {S}chmidt numbers},\ }\href
  {https://doi.org/https://doi.org/10.1016/0017-9310(75)90223-9} {\bibfield
  {journal} {\bibinfo  {journal} {Int. J. Heat Mass Transfer}\ }\textbf
  {\bibinfo {volume} {18}},\ \bibinfo {pages} {1055} (\bibinfo {year}
  {1975})}\BibitemShut {NoStop}%
\bibitem [{\citenamefont {Jischa}\ and\ \citenamefont
  {Rieke}(1979)}]{Jischa:IJHMT1979}%
  \BibitemOpen
  \bibfield  {author} {\bibinfo {author} {\bibfnamefont {M.}~\bibnamefont
  {Jischa}}\ and\ \bibinfo {author} {\bibfnamefont {H.~B.}\ \bibnamefont
  {Rieke}},\ }\bibfield  {title} {\bibinfo {title} {About the prediction of
  turbulent {P}randtl and {S}chmidt numbers from modeled transport equations},\
  }\href {https://doi.org/https://doi.org/10.1016/0017-9310(79)90134-0}
  {\bibfield  {journal} {\bibinfo  {journal} {Int. J. Heat Mass Transfer}\
  }\textbf {\bibinfo {volume} {22}},\ \bibinfo {pages} {1547} (\bibinfo {year}
  {1979})}\BibitemShut {NoStop}%
\bibitem [{\citenamefont {Tai}\ \emph {et~al.}(2021)\citenamefont {Tai},
  \citenamefont {Ching}, \citenamefont {Zwirner},\ and\ \citenamefont
  {Shishkina}}]{Tai:PRF2021}%
  \BibitemOpen
  \bibfield  {author} {\bibinfo {author} {\bibfnamefont {N.~C.}\ \bibnamefont
  {Tai}}, \bibinfo {author} {\bibfnamefont {E.~S.~C.}\ \bibnamefont {Ching}},
  \bibinfo {author} {\bibfnamefont {L.}~\bibnamefont {Zwirner}},\ and\ \bibinfo
  {author} {\bibfnamefont {O.}~\bibnamefont {Shishkina}},\ }\bibfield  {title}
  {\bibinfo {title} {Heat flux in turbulent {R}ayleigh-{B}\'enard convection:
  Predictions derived from a boundary layer theory},\ }\href
  {https://doi.org/10.1103/PhysRevFluids.6.033501} {\bibfield  {journal}
  {\bibinfo  {journal} {Phys. Rev. Fluids}\ }\textbf {\bibinfo {volume} {6}},\
  \bibinfo {pages} {033501} (\bibinfo {year} {2021})}\BibitemShut {NoStop}%
\bibitem [{\citenamefont {Horanyi}\ \emph {et~al.}(1999)\citenamefont
  {Horanyi}, \citenamefont {Krebs},\ and\ \citenamefont
  {M\"uller}}]{Horanyi:IJHMT1999}%
  \BibitemOpen
  \bibfield  {author} {\bibinfo {author} {\bibfnamefont {S.}~\bibnamefont
  {Horanyi}}, \bibinfo {author} {\bibfnamefont {L.}~\bibnamefont {Krebs}},\
  and\ \bibinfo {author} {\bibfnamefont {U.}~\bibnamefont {M\"uller}},\
  }\bibfield  {title} {\bibinfo {title} {Turbulent {R}ayleigh–{B}\'enard
  convection in low {P}randtl–number fluids},\ }\href
  {https://doi.org/https://doi.org/10.1016/S0017-9310(99)00059-9} {\bibfield
  {journal} {\bibinfo  {journal} {Int. J. Heat Mass Transfer}\ }\textbf
  {\bibinfo {volume} {42}},\ \bibinfo {pages} {3983} (\bibinfo {year}
  {1999})}\BibitemShut {NoStop}%
\bibitem [{\citenamefont {{Stevens}}\ \emph {et~al.}(2010)\citenamefont
  {{Stevens}}, \citenamefont {{Verzicco}},\ and\ \citenamefont
  {{Lohse}}}]{Stevens:JFM2010}%
  \BibitemOpen
  \bibfield  {author} {\bibinfo {author} {\bibfnamefont {R.}~\bibnamefont
  {{Stevens}}}, \bibinfo {author} {\bibfnamefont {R.}~\bibnamefont
  {{Verzicco}}},\ and\ \bibinfo {author} {\bibfnamefont {D.}~\bibnamefont
  {{Lohse}}},\ }\bibfield  {title} {\bibinfo {title} {Radial boundary layer
  structure and {N}usselt number in {R}ayleigh-{B}{\'e}nard convection},\
  }\href {https://doi.org/10.1017/S0022112009992461} {\bibfield  {journal}
  {\bibinfo  {journal} {J. Fluid Mech.}\ }\textbf {\bibinfo {volume} {643}},\
  \bibinfo {pages} {495} (\bibinfo {year} {2010})}\BibitemShut {NoStop}%
\bibitem [{\citenamefont {{Schumacher}}\ \emph {et~al.}(2015)\citenamefont
  {{Schumacher}}, \citenamefont {{G\"{o}tzfried}},\ and\ \citenamefont
  {{Scheel}}}]{Schumacher:PNAS2015}%
  \BibitemOpen
  \bibfield  {author} {\bibinfo {author} {\bibfnamefont {J.}~\bibnamefont
  {{Schumacher}}}, \bibinfo {author} {\bibfnamefont {P.}~\bibnamefont
  {{G\"{o}tzfried}}},\ and\ \bibinfo {author} {\bibfnamefont {J.~D.}\
  \bibnamefont {{Scheel}}},\ }\bibfield  {title} {\bibinfo {title} {Enhanced
  enstrophy generation for turbulent convection in low-{P}randtl-number
  fluids},\ }\href {https://doi.org/10.1073/pnas.1505111112} {\bibfield
  {journal} {\bibinfo  {journal} {Proc. Natl. Acad. Sci. USA}\ }\textbf
  {\bibinfo {volume} {112}},\ \bibinfo {pages} {9530} (\bibinfo {year}
  {2015})}\BibitemShut {NoStop}%
\bibitem [{\citenamefont {Pandey}\ and\ \citenamefont
  {Verma}(2016)}]{Pandey:POF2016}%
  \BibitemOpen
  \bibfield  {author} {\bibinfo {author} {\bibfnamefont {A.}~\bibnamefont
  {Pandey}}\ and\ \bibinfo {author} {\bibfnamefont {M.~K.}\ \bibnamefont
  {Verma}},\ }\bibfield  {title} {\bibinfo {title} {Scaling of large-scale
  quantities in {R}ayleigh-{B}{\'e}nard convection},\ }\href
  {https://doi.org/10.1063/1.4962307} {\bibfield  {journal} {\bibinfo
  {journal} {Phys. Fluids}\ }\textbf {\bibinfo {volume} {28}},\ \bibinfo
  {pages} {095105} (\bibinfo {year} {2016})},\ \Eprint
  {https://arxiv.org/abs/https://doi.org/10.1063/1.4962307}
  {https://doi.org/10.1063/1.4962307} \BibitemShut {NoStop}%
\bibitem [{\citenamefont {Scheel}\ and\ \citenamefont
  {Schumacher}(2017)}]{Scheel:PRF2017}%
  \BibitemOpen
  \bibfield  {author} {\bibinfo {author} {\bibfnamefont {J.~D.}\ \bibnamefont
  {Scheel}}\ and\ \bibinfo {author} {\bibfnamefont {J.}~\bibnamefont
  {Schumacher}},\ }\bibfield  {title} {\bibinfo {title} {Predicting transition
  ranges to fully turbulent viscous boundary layers in low {P}randtl number
  convection flows},\ }\href {https://doi.org/10.1103/PhysRevFluids.2.123501}
  {\bibfield  {journal} {\bibinfo  {journal} {Phys. Rev. Fluids}\ }\textbf
  {\bibinfo {volume} {2}},\ \bibinfo {pages} {123501} (\bibinfo {year}
  {2017})}\BibitemShut {NoStop}%
\bibitem [{\citenamefont {Pandey}\ \emph
  {et~al.}(2018{\natexlab{a}})\citenamefont {Pandey}, \citenamefont {Scheel},\
  and\ \citenamefont {Schumacher}}]{Pandey:Nature2018}%
  \BibitemOpen
  \bibfield  {author} {\bibinfo {author} {\bibfnamefont {A.}~\bibnamefont
  {Pandey}}, \bibinfo {author} {\bibfnamefont {J.~D.}\ \bibnamefont {Scheel}},\
  and\ \bibinfo {author} {\bibfnamefont {J.}~\bibnamefont {Schumacher}},\
  }\bibfield  {title} {\bibinfo {title} {Turbulent superstructures in
  {R}ayleigh-{B}\'enard convection},\ }\href
  {https://doi.org/10.1038/s41467-018-04478-0} {\bibfield  {journal} {\bibinfo
  {journal} {Nat. Commun.}\ }\textbf {\bibinfo {volume} {9}},\ \bibinfo {pages}
  {2118} (\bibinfo {year} {2018}{\natexlab{a}})}\BibitemShut {NoStop}%
\bibitem [{\citenamefont {Iyer}\ \emph {et~al.}(2020)\citenamefont {Iyer},
  \citenamefont {Scheel}, \citenamefont {Schumacher},\ and\ \citenamefont
  {Sreenivasan}}]{Iyer:PNAS2020}%
  \BibitemOpen
  \bibfield  {author} {\bibinfo {author} {\bibfnamefont {K.~P.}\ \bibnamefont
  {Iyer}}, \bibinfo {author} {\bibfnamefont {J.~D.}\ \bibnamefont {Scheel}},
  \bibinfo {author} {\bibfnamefont {J.}~\bibnamefont {Schumacher}},\ and\
  \bibinfo {author} {\bibfnamefont {K.~R.}\ \bibnamefont {Sreenivasan}},\
  }\bibfield  {title} {\bibinfo {title} {Classical 1/3 scaling of convection
  holds up to {R}a = $10^{15}$},\ }\href
  {https://doi.org/10.1073/pnas.1922794117} {\bibfield  {journal} {\bibinfo
  {journal} {Proc. Natl. Acad. Sci. USA}\ }\textbf {\bibinfo {volume} {117}},\
  \bibinfo {pages} {7594} (\bibinfo {year} {2020})},\ \Eprint
  {https://arxiv.org/abs/https://www.pnas.org/content/117/14/7594.full.pdf}
  {https://www.pnas.org/content/117/14/7594.full.pdf} \BibitemShut {NoStop}%
\bibitem [{\citenamefont {Sugiyama}\ \emph {et~al.}(2010)\citenamefont
  {Sugiyama}, \citenamefont {Ni}, \citenamefont {Stevens}, \citenamefont
  {Chan}, \citenamefont {Zhou}, \citenamefont {Xi}, \citenamefont {Sun},
  \citenamefont {Grossmann}, \citenamefont {Xia},\ and\ \citenamefont
  {Lohse}}]{Sugiyama:PRL2010}%
  \BibitemOpen
  \bibfield  {author} {\bibinfo {author} {\bibfnamefont {K.}~\bibnamefont
  {Sugiyama}}, \bibinfo {author} {\bibfnamefont {R.}~\bibnamefont {Ni}},
  \bibinfo {author} {\bibfnamefont {R.~J. A.~M.}\ \bibnamefont {Stevens}},
  \bibinfo {author} {\bibfnamefont {T.~S.}\ \bibnamefont {Chan}}, \bibinfo
  {author} {\bibfnamefont {S.-Q.}\ \bibnamefont {Zhou}}, \bibinfo {author}
  {\bibfnamefont {H.-D.}\ \bibnamefont {Xi}}, \bibinfo {author} {\bibfnamefont
  {C.}~\bibnamefont {Sun}}, \bibinfo {author} {\bibfnamefont {S.}~\bibnamefont
  {Grossmann}}, \bibinfo {author} {\bibfnamefont {K.-Q.}\ \bibnamefont {Xia}},\
  and\ \bibinfo {author} {\bibfnamefont {D.}~\bibnamefont {Lohse}},\ }\bibfield
   {title} {\bibinfo {title} {Flow reversals in thermally driven turbulence},\
  }\href {https://doi.org/10.1103/PhysRevLett.105.034503} {\bibfield  {journal}
  {\bibinfo  {journal} {Phys. Rev. Lett.}\ }\textbf {\bibinfo {volume} {105}},\
  \bibinfo {pages} {034503} (\bibinfo {year} {2010})}\BibitemShut {NoStop}%
\bibitem [{\citenamefont {Chandra}\ and\ \citenamefont
  {Verma}(2013)}]{Chandra:PRL2013}%
  \BibitemOpen
  \bibfield  {author} {\bibinfo {author} {\bibfnamefont {M.}~\bibnamefont
  {Chandra}}\ and\ \bibinfo {author} {\bibfnamefont {M.~K.}\ \bibnamefont
  {Verma}},\ }\bibfield  {title} {\bibinfo {title} {Flow reversals in turbulent
  convection via vortex reconnections},\ }\href
  {https://doi.org/10.1103/PhysRevLett.110.114503} {\bibfield  {journal}
  {\bibinfo  {journal} {Phys. Rev. Lett.}\ }\textbf {\bibinfo {volume} {110}},\
  \bibinfo {pages} {114503} (\bibinfo {year} {2013})}\BibitemShut {NoStop}%
\bibitem [{\citenamefont {{Podvin}}\ and\ \citenamefont
  {{Sergent}}(2015)}]{Podvin:JFM2015}%
  \BibitemOpen
  \bibfield  {author} {\bibinfo {author} {\bibfnamefont {B.}~\bibnamefont
  {{Podvin}}}\ and\ \bibinfo {author} {\bibfnamefont {A.}~\bibnamefont
  {{Sergent}}},\ }\bibfield  {title} {\bibinfo {title} {A large-scale
  investigation of wind reversal in a square {R}ayleigh-{B}{\'e}nard cell},\
  }\href {https://doi.org/10.1017/jfm.2015.15} {\bibfield  {journal} {\bibinfo
  {journal} {J. Fluid Mech.}\ }\textbf {\bibinfo {volume} {766}},\ \bibinfo
  {pages} {172} (\bibinfo {year} {2015})}\BibitemShut {NoStop}%
\bibitem [{\citenamefont {Pandey}\ \emph
  {et~al.}(2018{\natexlab{b}})\citenamefont {Pandey}, \citenamefont {Verma},\
  and\ \citenamefont {Barma}}]{Pandey:PRE2018}%
  \BibitemOpen
  \bibfield  {author} {\bibinfo {author} {\bibfnamefont {A.}~\bibnamefont
  {Pandey}}, \bibinfo {author} {\bibfnamefont {M.~K.}\ \bibnamefont {Verma}},\
  and\ \bibinfo {author} {\bibfnamefont {M.}~\bibnamefont {Barma}},\ }\bibfield
   {title} {\bibinfo {title} {Reversals in infinite-{P}randtl-number
  {R}ayleigh-{B}\'enard convection},\ }\href
  {https://doi.org/10.1103/PhysRevE.98.023109} {\bibfield  {journal} {\bibinfo
  {journal} {Phys. Rev. E}\ }\textbf {\bibinfo {volume} {98}},\ \bibinfo
  {pages} {023109} (\bibinfo {year} {2018}{\natexlab{b}})}\BibitemShut
  {NoStop}%
\bibitem [{\citenamefont {Zhu}\ \emph {et~al.}(2018)\citenamefont {Zhu},
  \citenamefont {Mathai}, \citenamefont {Stevens}, \citenamefont {Verzicco},\
  and\ \citenamefont {Lohse}}]{Zhu:PRL2018}%
  \BibitemOpen
  \bibfield  {author} {\bibinfo {author} {\bibfnamefont {X.}~\bibnamefont
  {Zhu}}, \bibinfo {author} {\bibfnamefont {V.}~\bibnamefont {Mathai}},
  \bibinfo {author} {\bibfnamefont {R.~J. A.~M.}\ \bibnamefont {Stevens}},
  \bibinfo {author} {\bibfnamefont {R.}~\bibnamefont {Verzicco}},\ and\
  \bibinfo {author} {\bibfnamefont {D.}~\bibnamefont {Lohse}},\ }\bibfield
  {title} {\bibinfo {title} {Transition to the ultimate regime in
  two-dimensional {R}ayleigh-{B}\'enard convection},\ }\href
  {https://doi.org/10.1103/PhysRevLett.120.144502} {\bibfield  {journal}
  {\bibinfo  {journal} {Phys. Rev. Lett.}\ }\textbf {\bibinfo {volume} {120}},\
  \bibinfo {pages} {144502} (\bibinfo {year} {2018})}\BibitemShut {NoStop}%
\bibitem [{\citenamefont {{van der Poel}}\ \emph {et~al.}(2015)\citenamefont
  {{van der Poel}}, \citenamefont {{Ostilla-M\'{o}nico}}, \citenamefont
  {{Verzicco}}, \citenamefont {{Grossmann}},\ and\ \citenamefont
  {{Lohse}}}]{Poel:PRL2015}%
  \BibitemOpen
  \bibfield  {author} {\bibinfo {author} {\bibfnamefont {E.~P.}\ \bibnamefont
  {{van der Poel}}}, \bibinfo {author} {\bibfnamefont {R.}~\bibnamefont
  {{Ostilla-M\'{o}nico}}}, \bibinfo {author} {\bibfnamefont {R.}~\bibnamefont
  {{Verzicco}}}, \bibinfo {author} {\bibfnamefont {S.}~\bibnamefont
  {{Grossmann}}},\ and\ \bibinfo {author} {\bibfnamefont {D.}~\bibnamefont
  {{Lohse}}},\ }\bibfield  {title} {\bibinfo {title} {Logarithmic mean
  temperature profiles and their connection to plume emissions in turbulent
  {R}ayleigh-{B}{\'e}nard convection},\ }\href
  {https://doi.org/10.1103/PhysRevLett.115.154501} {\bibfield  {journal}
  {\bibinfo  {journal} {Phys. Rev. Lett.}\ }\textbf {\bibinfo {volume} {115}},\
  \bibinfo {pages} {154501} (\bibinfo {year} {2015})}\BibitemShut {NoStop}%
\bibitem [{\citenamefont {Zhou}\ \emph {et~al.}(2011)\citenamefont {Zhou},
  \citenamefont {Sugiyama}, \citenamefont {Stevens}, \citenamefont {Grossmann},
  \citenamefont {Lohse},\ and\ \citenamefont {Xia}}]{Zhou:POF2011}%
  \BibitemOpen
  \bibfield  {author} {\bibinfo {author} {\bibfnamefont {Q.}~\bibnamefont
  {Zhou}}, \bibinfo {author} {\bibfnamefont {K.}~\bibnamefont {Sugiyama}},
  \bibinfo {author} {\bibfnamefont {R.~J. A.~M.}\ \bibnamefont {Stevens}},
  \bibinfo {author} {\bibfnamefont {S.}~\bibnamefont {Grossmann}}, \bibinfo
  {author} {\bibfnamefont {D.}~\bibnamefont {Lohse}},\ and\ \bibinfo {author}
  {\bibfnamefont {K.-Q.}\ \bibnamefont {Xia}},\ }\bibfield  {title} {\bibinfo
  {title} {Horizontal structures of velocity and temperature boundary layers in
  two-dimensional numerical turbulent {R}ayleigh-{B}{\'e}nard convection},\
  }\href {https://doi.org/10.1063/1.3662445} {\bibfield  {journal} {\bibinfo
  {journal} {Phys. Fluids}\ }\textbf {\bibinfo {volume} {23}},\ \bibinfo
  {pages} {125104} (\bibinfo {year} {2011})},\ \Eprint
  {https://arxiv.org/abs/https://doi.org/10.1063/1.3662445}
  {https://doi.org/10.1063/1.3662445} \BibitemShut {NoStop}%
\bibitem [{\citenamefont {Pandey}(2021)}]{Pandey:JFM2021}%
  \BibitemOpen
  \bibfield  {author} {\bibinfo {author} {\bibfnamefont {A.}~\bibnamefont
  {Pandey}},\ }\bibfield  {title} {\bibinfo {title} {Thermal boundary layer
  structure in low-{P}randtl-number turbulent convection},\ }\href
  {https://doi.org/10.1017/jfm.2020.961} {\bibfield  {journal} {\bibinfo
  {journal} {J. Fluid Mech.}\ }\textbf {\bibinfo {volume} {910}},\ \bibinfo
  {pages} {A13} (\bibinfo {year} {2021})}\BibitemShut {NoStop}%
\bibitem [{\citenamefont {{Chandrasekhar}}(1981)}]{Chandrasekhar:book}%
  \BibitemOpen
  \bibfield  {author} {\bibinfo {author} {\bibfnamefont {S.}~\bibnamefont
  {{Chandrasekhar}}},\ }\href@noop {} {\emph {\bibinfo {title} {Hydrodynamic
  and Hydromagnetic Stability}}}\ (\bibinfo  {publisher} {Dover},\ \bibinfo
  {address} {New York},\ \bibinfo {year} {1981})\BibitemShut {NoStop}%
\bibitem [{\citenamefont {Zhang}\ \emph {et~al.}(1997)\citenamefont {Zhang},
  \citenamefont {Childress},\ and\ \citenamefont {Libchaber}}]{Zhang:POF1997}%
  \BibitemOpen
  \bibfield  {author} {\bibinfo {author} {\bibfnamefont {J.}~\bibnamefont
  {Zhang}}, \bibinfo {author} {\bibfnamefont {S.}~\bibnamefont {Childress}},\
  and\ \bibinfo {author} {\bibfnamefont {A.}~\bibnamefont {Libchaber}},\
  }\bibfield  {title} {\bibinfo {title} {Non-{B}oussinesq effect: Thermal
  convection with broken symmetry},\ }\href {https://doi.org/10.1063/1.869198}
  {\bibfield  {journal} {\bibinfo  {journal} {Phys. Fluids}\ }\textbf {\bibinfo
  {volume} {9}},\ \bibinfo {pages} {1034} (\bibinfo {year} {1997})},\ \Eprint
  {https://arxiv.org/abs/https://doi.org/10.1063/1.869198}
  {https://doi.org/10.1063/1.869198} \BibitemShut {NoStop}%
\bibitem [{\citenamefont {{Ahlers}}\ \emph {et~al.}(2006)\citenamefont
  {{Ahlers}}, \citenamefont {{Brown}}, \citenamefont {{Araujo}}, \citenamefont
  {{Funfschilling}}, \citenamefont {{Grossmann}},\ and\ \citenamefont
  {{Lohse}}}]{Ahlers:JFM2006}%
  \BibitemOpen
  \bibfield  {author} {\bibinfo {author} {\bibfnamefont {G.}~\bibnamefont
  {{Ahlers}}}, \bibinfo {author} {\bibfnamefont {E.}~\bibnamefont {{Brown}}},
  \bibinfo {author} {\bibfnamefont {F.~F.}\ \bibnamefont {{Araujo}}}, \bibinfo
  {author} {\bibfnamefont {D.}~\bibnamefont {{Funfschilling}}}, \bibinfo
  {author} {\bibfnamefont {S.}~\bibnamefont {{Grossmann}}},\ and\ \bibinfo
  {author} {\bibfnamefont {D.}~\bibnamefont {{Lohse}}},\ }\bibfield  {title}
  {\bibinfo {title} {Non-{O}berbeck-{B}oussinesq effects in strongly turbulent
  {R}ayleigh-{B}{\'e}nard convection},\ }\href
  {https://doi.org/10.1017/S0022112006002916} {\bibfield  {journal} {\bibinfo
  {journal} {J. Fluid Mech.}\ }\textbf {\bibinfo {volume} {569}},\ \bibinfo
  {pages} {409} (\bibinfo {year} {2006})}\BibitemShut {NoStop}%
\bibitem [{\citenamefont {Sameen}\ \emph {et~al.}(2008)\citenamefont {Sameen},
  \citenamefont {Verzicco},\ and\ \citenamefont
  {Sreenivasan}}]{Sameen:PhSc2008}%
  \BibitemOpen
  \bibfield  {author} {\bibinfo {author} {\bibfnamefont {A.}~\bibnamefont
  {Sameen}}, \bibinfo {author} {\bibfnamefont {R.}~\bibnamefont {Verzicco}},\
  and\ \bibinfo {author} {\bibfnamefont {K.~R.}\ \bibnamefont {Sreenivasan}},\
  }\bibfield  {title} {\bibinfo {title} {Non-{B}oussinesq convection at
  moderate {R}ayleigh numbers in low temperature gaseous helium},\ }\href
  {https://doi.org/10.1088/0031-8949/2008/t132/014053} {\bibfield  {journal}
  {\bibinfo  {journal} {Phys. Scr.}\ }\textbf {\bibinfo {volume} {T132}},\
  \bibinfo {pages} {014053} (\bibinfo {year} {2008})}\BibitemShut {NoStop}%
\bibitem [{\citenamefont {Sameen}\ \emph {et~al.}(2009)\citenamefont {Sameen},
  \citenamefont {Verzicco},\ and\ \citenamefont
  {Sreenivasan}}]{Sameen:EPL2009}%
  \BibitemOpen
  \bibfield  {author} {\bibinfo {author} {\bibfnamefont {A.}~\bibnamefont
  {Sameen}}, \bibinfo {author} {\bibfnamefont {R.}~\bibnamefont {Verzicco}},\
  and\ \bibinfo {author} {\bibfnamefont {K.~R.}\ \bibnamefont {Sreenivasan}},\
  }\bibfield  {title} {\bibinfo {title} {Specific roles of fluid properties in
  non-{B}oussinesq thermal convection at the {R}ayleigh number of
  $2\times10^8$},\ }\href {https://doi.org/10.1209/0295-5075/86/14006}
  {\bibfield  {journal} {\bibinfo  {journal} {Europhys. Lett.}\ }\textbf
  {\bibinfo {volume} {86}},\ \bibinfo {pages} {14006} (\bibinfo {year}
  {2009})}\BibitemShut {NoStop}%
\bibitem [{\citenamefont {{Sugiyama}}\ \emph {et~al.}(2009)\citenamefont
  {{Sugiyama}}, \citenamefont {{Calzavarini}}, \citenamefont {{Grossmann}},\
  and\ \citenamefont {{Lohse}}}]{Sugiyama:JFM2009}%
  \BibitemOpen
  \bibfield  {author} {\bibinfo {author} {\bibfnamefont {K.}~\bibnamefont
  {{Sugiyama}}}, \bibinfo {author} {\bibfnamefont {E.}~\bibnamefont
  {{Calzavarini}}}, \bibinfo {author} {\bibfnamefont {S.}~\bibnamefont
  {{Grossmann}}},\ and\ \bibinfo {author} {\bibfnamefont {D.}~\bibnamefont
  {{Lohse}}},\ }\bibfield  {title} {\bibinfo {title} {Flow organization in
  two-dimensional non-{O}berbeck–{B}oussinesq {R}ayleigh–{B}\'enard
  convection in water},\ }\href {https://doi.org/10.1017/S0022112009008027}
  {\bibfield  {journal} {\bibinfo  {journal} {J. Fluid Mech.}\ }\textbf
  {\bibinfo {volume} {637}},\ \bibinfo {pages} {105–135} (\bibinfo {year}
  {2009})}\BibitemShut {NoStop}%
\bibitem [{\citenamefont {{Horn}}\ \emph {et~al.}(2013)\citenamefont {{Horn}},
  \citenamefont {{Shishkina}},\ and\ \citenamefont {{Wagner}}}]{Horn:JFM2013}%
  \BibitemOpen
  \bibfield  {author} {\bibinfo {author} {\bibfnamefont {S.}~\bibnamefont
  {{Horn}}}, \bibinfo {author} {\bibfnamefont {O.}~\bibnamefont
  {{Shishkina}}},\ and\ \bibinfo {author} {\bibfnamefont {C.}~\bibnamefont
  {{Wagner}}},\ }\bibfield  {title} {\bibinfo {title} {On
  non-{O}berbeck-{B}oussinesq effects in three-dimensional
  {R}ayleigh-{B}{\'e}nard convection in glycerol},\ }\href
  {https://doi.org/10.1017/jfm.2013.151} {\bibfield  {journal} {\bibinfo
  {journal} {J. Fluid Mech.}\ }\textbf {\bibinfo {volume} {724}},\ \bibinfo
  {pages} {175} (\bibinfo {year} {2013})}\BibitemShut {NoStop}%
\bibitem [{\citenamefont {Tritton}(1977)}]{Tritton:book1977}%
  \BibitemOpen
  \bibfield  {author} {\bibinfo {author} {\bibfnamefont {D.~J.}\ \bibnamefont
  {Tritton}},\ }\href {https://doi.org/10.1007/978-94-009-9992-3} {\emph
  {\bibinfo {title} {Physical Fluid Dynamics}}}\ (\bibinfo  {publisher}
  {Springer Netherlands},\ \bibinfo {address} {Dordrecht},\ \bibinfo {year}
  {1977})\ \Eprint
  {https://arxiv.org/abs/https://www.springer.com/gp/book/9780442301323}
  {https://www.springer.com/gp/book/9780442301323} \BibitemShut {NoStop}%
\bibitem [{\citenamefont {Shcheritsa}\ \emph {et~al.}(2018)\citenamefont
  {Shcheritsa}, \citenamefont {Getling},\ and\ \citenamefont
  {Mazhorova}}]{Getling:PLA2018}%
  \BibitemOpen
  \bibfield  {author} {\bibinfo {author} {\bibfnamefont {O.}~\bibnamefont
  {Shcheritsa}}, \bibinfo {author} {\bibfnamefont {A.}~\bibnamefont
  {Getling}},\ and\ \bibinfo {author} {\bibfnamefont {O.}~\bibnamefont
  {Mazhorova}},\ }\bibfield  {title} {\bibinfo {title} {Effects of variable
  thermal diffusivity on the structure of convection},\ }\href
  {https://doi.org/https://doi.org/10.1016/j.physleta.2018.01.009} {\bibfield
  {journal} {\bibinfo  {journal} {Phys. Lett. A}\ }\textbf {\bibinfo {volume}
  {382}},\ \bibinfo {pages} {639 } (\bibinfo {year} {2018})}\BibitemShut
  {NoStop}%
\bibitem [{\citenamefont {Pandey}\ \emph {et~al.}(2021)\citenamefont {Pandey},
  \citenamefont {Schumacher},\ and\ \citenamefont
  {Sreenivasan}}]{Pandey:APJ2021}%
  \BibitemOpen
  \bibfield  {author} {\bibinfo {author} {\bibfnamefont {A.}~\bibnamefont
  {Pandey}}, \bibinfo {author} {\bibfnamefont {J.}~\bibnamefont {Schumacher}},\
  and\ \bibinfo {author} {\bibfnamefont {K.~R.}\ \bibnamefont {Sreenivasan}},\
  }\bibfield  {title} {\bibinfo {title} {Non-{B}oussinesq low-{P}randtl-number
  convection with a temperature-dependent thermal diffusivity},\ }\href
  {https://doi.org/10.3847/1538-4357/abd1d8} {\bibfield  {journal} {\bibinfo
  {journal} {Astrophys. J.}\ }\textbf {\bibinfo {volume} {907}},\ \bibinfo
  {pages} {56} (\bibinfo {year} {2021})}\BibitemShut {NoStop}%
\bibitem [{\citenamefont {{Fischer}}(1997)}]{Fischer:JCP1997}%
  \BibitemOpen
  \bibfield  {author} {\bibinfo {author} {\bibfnamefont {P.~F.}\ \bibnamefont
  {{Fischer}}},\ }\bibfield  {title} {\bibinfo {title} {An overlapping
  {S}chwarz method for spectral element solution of the incompressible
  {N}avier-{S}tokes equations},\ }\href
  {https://doi.org/10.1006/jcph.1997.5651} {\bibfield  {journal} {\bibinfo
  {journal} {J. Comp. Phys.}\ }\textbf {\bibinfo {volume} {133}},\ \bibinfo
  {pages} {84} (\bibinfo {year} {1997})}\BibitemShut {NoStop}%
\bibitem [{\citenamefont {{Scheel}}\ \emph {et~al.}(2013)\citenamefont
  {{Scheel}}, \citenamefont {{Emran}},\ and\ \citenamefont
  {{Schumacher}}}]{Scheel:NJP2013}%
  \BibitemOpen
  \bibfield  {author} {\bibinfo {author} {\bibfnamefont {J.~D.}\ \bibnamefont
  {{Scheel}}}, \bibinfo {author} {\bibfnamefont {M.~S.}\ \bibnamefont
  {{Emran}}},\ and\ \bibinfo {author} {\bibfnamefont {J.}~\bibnamefont
  {{Schumacher}}},\ }\bibfield  {title} {\bibinfo {title} {Resolving the
  fine-scale structure in turbulent {R}ayleigh-{B}{\'e}nard convection},\
  }\href {https://doi.org/10.1088/1367-2630/15/11/113063} {\bibfield  {journal}
  {\bibinfo  {journal} {New J. Phys.}\ }\textbf {\bibinfo {volume} {15}},\
  \bibinfo {pages} {113063} (\bibinfo {year} {2013})}\BibitemShut {NoStop}%
\bibitem [{\citenamefont {Pandey}\ and\ \citenamefont
  {Sreenivasan}(2021)}]{Pandey:EPL2021}%
  \BibitemOpen
  \bibfield  {author} {\bibinfo {author} {\bibfnamefont {A.}~\bibnamefont
  {Pandey}}\ and\ \bibinfo {author} {\bibfnamefont {K.~R.}\ \bibnamefont
  {Sreenivasan}},\ }\bibfield  {title} {\bibinfo {title} {Convective heat
  transport in slender cells is close to that in wider cells at high {R}ayleigh
  and {P}randtl numbers},\ }\href {https://doi.org/10.1209/0295-5075/ac1bc9}
  {\bibfield  {journal} {\bibinfo  {journal} {Europhys. Lett.}\ }\textbf
  {\bibinfo {volume} {135}},\ \bibinfo {pages} {24001} (\bibinfo {year}
  {2021})}\BibitemShut {NoStop}%
\bibitem [{\citenamefont {{Silano}}\ \emph {et~al.}(2010)\citenamefont
  {{Silano}}, \citenamefont {{Sreenivasan}},\ and\ \citenamefont
  {{Verzicco}}}]{Silano:JFM2010}%
  \BibitemOpen
  \bibfield  {author} {\bibinfo {author} {\bibfnamefont {G.}~\bibnamefont
  {{Silano}}}, \bibinfo {author} {\bibfnamefont {K.~R.}\ \bibnamefont
  {{Sreenivasan}}},\ and\ \bibinfo {author} {\bibfnamefont {R.}~\bibnamefont
  {{Verzicco}}},\ }\bibfield  {title} {\bibinfo {title} {Numerical simulations
  of {R}ayleigh-{B}{\'e}nard convection for {P}randtl numbers between $10^{-1}$
  and $10^4$ and {R}ayleigh numbers between $10^5$ and $10^9$},\ }\href
  {https://doi.org/10.1017/S0022112010003290} {\bibfield  {journal} {\bibinfo
  {journal} {J. Fluid Mech.}\ }\textbf {\bibinfo {volume} {662}},\ \bibinfo
  {pages} {409} (\bibinfo {year} {2010})}\BibitemShut {NoStop}%
\bibitem [{\citenamefont {Pandey}\ \emph {et~al.}(2014)\citenamefont {Pandey},
  \citenamefont {Verma},\ and\ \citenamefont {Mishra}}]{Pandey:PRE2014}%
  \BibitemOpen
  \bibfield  {author} {\bibinfo {author} {\bibfnamefont {A.}~\bibnamefont
  {Pandey}}, \bibinfo {author} {\bibfnamefont {M.~K.}\ \bibnamefont {Verma}},\
  and\ \bibinfo {author} {\bibfnamefont {P.~K.}\ \bibnamefont {Mishra}},\
  }\bibfield  {title} {\bibinfo {title} {Scaling of heat flux and energy
  spectrum for very large {P}randtl number convection},\ }\href
  {https://doi.org/10.1103/PhysRevE.89.023006} {\bibfield  {journal} {\bibinfo
  {journal} {Phys. Rev. E}\ }\textbf {\bibinfo {volume} {89}},\ \bibinfo
  {pages} {023006} (\bibinfo {year} {2014})}\BibitemShut {NoStop}%
\bibitem [{\citenamefont {{Pandey}}\ \emph {et~al.}(2016)\citenamefont
  {{Pandey}}, \citenamefont {{Verma}}, \citenamefont {{Chatterjee}},\ and\
  \citenamefont {{Dutta}}}]{Pandey:Pramana2016}%
  \BibitemOpen
  \bibfield  {author} {\bibinfo {author} {\bibfnamefont {A.}~\bibnamefont
  {{Pandey}}}, \bibinfo {author} {\bibfnamefont {M.~K.}\ \bibnamefont
  {{Verma}}}, \bibinfo {author} {\bibfnamefont {A.~G.}\ \bibnamefont
  {{Chatterjee}}},\ and\ \bibinfo {author} {\bibfnamefont {B.}~\bibnamefont
  {{Dutta}}},\ }\bibfield  {title} {\bibinfo {title} {Similarities between 2{D}
  and 3{D} convection for large {P}randtl number},\ }\href
  {https://doi.org/10.1007/s12043-016-1204-z} {\bibfield  {journal} {\bibinfo
  {journal} {Pramana - J. Phys.}\ }\textbf {\bibinfo {volume} {87}},\ \bibinfo
  {pages} {13} (\bibinfo {year} {2016})}\BibitemShut {NoStop}%
\bibitem [{\citenamefont {Shraiman}\ and\ \citenamefont
  {Siggia}(1990)}]{Shraiman:PRA1990}%
  \BibitemOpen
  \bibfield  {author} {\bibinfo {author} {\bibfnamefont {B.~I.}\ \bibnamefont
  {Shraiman}}\ and\ \bibinfo {author} {\bibfnamefont {E.~D.}\ \bibnamefont
  {Siggia}},\ }\bibfield  {title} {\bibinfo {title} {Heat transport in
  high-{R}ayleigh-number convection},\ }\href
  {https://doi.org/10.1103/PhysRevA.42.3650} {\bibfield  {journal} {\bibinfo
  {journal} {Phys. Rev. A}\ }\textbf {\bibinfo {volume} {42}},\ \bibinfo
  {pages} {3650} (\bibinfo {year} {1990})}\BibitemShut {NoStop}%
\bibitem [{\citenamefont {Jones}\ \emph {et~al.}(1976)\citenamefont {Jones},
  \citenamefont {Moore},\ and\ \citenamefont {Weiss}}]{Jones:JFM1976}%
  \BibitemOpen
  \bibfield  {author} {\bibinfo {author} {\bibfnamefont {C.~A.}\ \bibnamefont
  {Jones}}, \bibinfo {author} {\bibfnamefont {D.~R.}\ \bibnamefont {Moore}},\
  and\ \bibinfo {author} {\bibfnamefont {N.~O.}\ \bibnamefont {Weiss}},\
  }\bibfield  {title} {\bibinfo {title} {Axisymmetric convection in a
  cylinder},\ }\href {https://doi.org/10.1017/S0022112076001407} {\bibfield
  {journal} {\bibinfo  {journal} {J. Fluid Mech.}\ }\textbf {\bibinfo {volume}
  {73}},\ \bibinfo {pages} {353–388} (\bibinfo {year} {1976})}\BibitemShut
  {NoStop}%
\bibitem [{\citenamefont {Clever}\ and\ \citenamefont
  {Busse}(1981)}]{Clever:JFM1981}%
  \BibitemOpen
  \bibfield  {author} {\bibinfo {author} {\bibfnamefont {R.~M.}\ \bibnamefont
  {Clever}}\ and\ \bibinfo {author} {\bibfnamefont {F.~H.}\ \bibnamefont
  {Busse}},\ }\bibfield  {title} {\bibinfo {title} {Low-{P}randtl-number
  convection in a layer heated from below},\ }\href
  {https://doi.org/10.1017/S002211208100253X} {\bibfield  {journal} {\bibinfo
  {journal} {J. Fluid Mech.}\ }\textbf {\bibinfo {volume} {102}},\ \bibinfo
  {pages} {61–74} (\bibinfo {year} {1981})}\BibitemShut {NoStop}%
\bibitem [{\citenamefont {Busse}\ and\ \citenamefont
  {Clever}(1981)}]{Busse:JFM1981}%
  \BibitemOpen
  \bibfield  {author} {\bibinfo {author} {\bibfnamefont {F.~H.}\ \bibnamefont
  {Busse}}\ and\ \bibinfo {author} {\bibfnamefont {R.~M.}\ \bibnamefont
  {Clever}},\ }\bibfield  {title} {\bibinfo {title} {An asymptotic model of
  two-dimensional convection in the limit of low {P}randtl number},\ }\href
  {https://doi.org/10.1017/S0022112081002541} {\bibfield  {journal} {\bibinfo
  {journal} {J. Fluid Mech.}\ }\textbf {\bibinfo {volume} {102}},\ \bibinfo
  {pages} {75–83} (\bibinfo {year} {1981})}\BibitemShut {NoStop}%
\bibitem [{\citenamefont {Thual}(1992)}]{Thual:JFM1992}%
  \BibitemOpen
  \bibfield  {author} {\bibinfo {author} {\bibfnamefont {O.}~\bibnamefont
  {Thual}},\ }\bibfield  {title} {\bibinfo {title} {Zero-{P}randtl-number
  convection},\ }\href {https://doi.org/10.1017/S0022112092000089} {\bibfield
  {journal} {\bibinfo  {journal} {J. Fluid Mech.}\ }\textbf {\bibinfo {volume}
  {240}},\ \bibinfo {pages} {229–258} (\bibinfo {year} {1992})}\BibitemShut
  {NoStop}%
\bibitem [{\citenamefont {{Emran}}\ and\ \citenamefont
  {{Schumacher}}(2015)}]{Emran:JFM2015}%
  \BibitemOpen
  \bibfield  {author} {\bibinfo {author} {\bibfnamefont {M.~S.}\ \bibnamefont
  {{Emran}}}\ and\ \bibinfo {author} {\bibfnamefont {J.}~\bibnamefont
  {{Schumacher}}},\ }\bibfield  {title} {\bibinfo {title} {Large-scale mean
  patterns in turbulent convection},\ }\href
  {https://doi.org/10.1017/jfm.2015.316} {\bibfield  {journal} {\bibinfo
  {journal} {J. Fluid Mech.}\ }\textbf {\bibinfo {volume} {776}},\ \bibinfo
  {pages} {96} (\bibinfo {year} {2015})}\BibitemShut {NoStop}%
\bibitem [{\citenamefont {Shishkina}\ \emph {et~al.}(2017)\citenamefont
  {Shishkina}, \citenamefont {Horn}, \citenamefont {Emran},\ and\ \citenamefont
  {Ching}}]{Shishkina:PRF2017}%
  \BibitemOpen
  \bibfield  {author} {\bibinfo {author} {\bibfnamefont {O.}~\bibnamefont
  {Shishkina}}, \bibinfo {author} {\bibfnamefont {S.}~\bibnamefont {Horn}},
  \bibinfo {author} {\bibfnamefont {M.~S.}\ \bibnamefont {Emran}},\ and\
  \bibinfo {author} {\bibfnamefont {E.~S.~C.}\ \bibnamefont {Ching}},\
  }\bibfield  {title} {\bibinfo {title} {Mean temperature profiles in turbulent
  thermal convection},\ }\href {https://doi.org/10.1103/PhysRevFluids.2.113502}
  {\bibfield  {journal} {\bibinfo  {journal} {Phys. Rev. Fluids}\ }\textbf
  {\bibinfo {volume} {2}},\ \bibinfo {pages} {113502} (\bibinfo {year}
  {2017})}\BibitemShut {NoStop}%
\bibitem [{\citenamefont {Ching}\ \emph {et~al.}(2019)\citenamefont {Ching},
  \citenamefont {Leung}, \citenamefont {Zwirner},\ and\ \citenamefont
  {Shishkina}}]{Ching:PRR2019}%
  \BibitemOpen
  \bibfield  {author} {\bibinfo {author} {\bibfnamefont {E.~S.~C.}\
  \bibnamefont {Ching}}, \bibinfo {author} {\bibfnamefont {H.~S.}\ \bibnamefont
  {Leung}}, \bibinfo {author} {\bibfnamefont {L.}~\bibnamefont {Zwirner}},\
  and\ \bibinfo {author} {\bibfnamefont {O.}~\bibnamefont {Shishkina}},\
  }\bibfield  {title} {\bibinfo {title} {Velocity and thermal boundary layer
  equations for turbulent {R}ayleigh-{B}\'enard convection},\ }\href
  {https://doi.org/10.1103/PhysRevResearch.1.033037} {\bibfield  {journal}
  {\bibinfo  {journal} {Phys. Rev. Research}\ }\textbf {\bibinfo {volume}
  {1}},\ \bibinfo {pages} {033037} (\bibinfo {year} {2019})}\BibitemShut
  {NoStop}%
\bibitem [{\citenamefont {Davidson}(2004)}]{Davidson:book2004}%
  \BibitemOpen
  \bibfield  {author} {\bibinfo {author} {\bibfnamefont {P.~A.}\ \bibnamefont
  {Davidson}},\ }\href@noop {} {\emph {\bibinfo {title} {Turbulence: an
  introduction for scientists and engineers}}}\ (\bibinfo  {publisher} {Oxford
  University Press},\ \bibinfo {address} {Oxford, UK},\ \bibinfo {year}
  {2004})\BibitemShut {NoStop}%
\bibitem [{\citenamefont {Yakhot}\ and\ \citenamefont
  {Orszag}(1986)}]{Yakhot:JSC1986}%
  \BibitemOpen
  \bibfield  {author} {\bibinfo {author} {\bibfnamefont {V.}~\bibnamefont
  {Yakhot}}\ and\ \bibinfo {author} {\bibfnamefont {S.~A.}\ \bibnamefont
  {Orszag}},\ }\bibfield  {title} {\bibinfo {title} {Renormalization group
  analysis of turbulence. {I}. basic theory},\ }\href
  {https://doi.org/10.1007/BF01061452} {\bibfield  {journal} {\bibinfo
  {journal} {J. Sci. Comp.}\ }\textbf {\bibinfo {volume} {1}},\ \bibinfo
  {pages} {3} (\bibinfo {year} {1986})}\BibitemShut {NoStop}%
\bibitem [{\citenamefont {Deardorff}\ and\ \citenamefont
  {Willis}(1967)}]{Deardorff:JFM1967}%
  \BibitemOpen
  \bibfield  {author} {\bibinfo {author} {\bibfnamefont {J.~W.}\ \bibnamefont
  {Deardorff}}\ and\ \bibinfo {author} {\bibfnamefont {G.~E.}\ \bibnamefont
  {Willis}},\ }\bibfield  {title} {\bibinfo {title} {Investigation of turbulent
  thermal convection between horizontal plates},\ }\href
  {https://doi.org/10.1017/S0022112067002393} {\bibfield  {journal} {\bibinfo
  {journal} {J. Fluid Mech.}\ }\textbf {\bibinfo {volume} {28}},\ \bibinfo
  {pages} {675–704} (\bibinfo {year} {1967})}\BibitemShut {NoStop}%
\bibitem [{\citenamefont {Adrian}(1996)}]{Adrian:IJHMT1996}%
  \BibitemOpen
  \bibfield  {author} {\bibinfo {author} {\bibfnamefont {R.~J.}\ \bibnamefont
  {Adrian}},\ }\bibfield  {title} {\bibinfo {title} {Variation of temperature
  and velocity fluctuations in turbulent thermal convection over horizontal
  surfaces},\ }\href
  {https://doi.org/https://doi.org/10.1016/0017-9310(95)00317-7} {\bibfield
  {journal} {\bibinfo  {journal} {Int. J. Heat Mass Transfer}\ }\textbf
  {\bibinfo {volume} {39}},\ \bibinfo {pages} {2303} (\bibinfo {year}
  {1996})}\BibitemShut {NoStop}%
\bibitem [{\citenamefont {Scheel}\ and\ \citenamefont
  {Schumacher}(2016)}]{Scheel:JFM2016}%
  \BibitemOpen
  \bibfield  {author} {\bibinfo {author} {\bibfnamefont {J.~D.}\ \bibnamefont
  {Scheel}}\ and\ \bibinfo {author} {\bibfnamefont {J.}~\bibnamefont
  {Schumacher}},\ }\bibfield  {title} {\bibinfo {title} {Global and local
  statistics in turbulent convection at low {P}randtl numbers},\ }\href
  {https://doi.org/10.1017/jfm.2016.457} {\bibfield  {journal} {\bibinfo
  {journal} {J. Fluid Mech.}\ }\textbf {\bibinfo {volume} {802}},\ \bibinfo
  {pages} {147–173} (\bibinfo {year} {2016})}\BibitemShut {NoStop}%
\bibitem [{\citenamefont {{Shishkina}}\ \emph {et~al.}(2015)\citenamefont
  {{Shishkina}}, \citenamefont {{Horn}}, \citenamefont {{Wagner}},\ and\
  \citenamefont {{Ching}}}]{Shishkina:PRL2015}%
  \BibitemOpen
  \bibfield  {author} {\bibinfo {author} {\bibfnamefont {O.}~\bibnamefont
  {{Shishkina}}}, \bibinfo {author} {\bibfnamefont {S.}~\bibnamefont {{Horn}}},
  \bibinfo {author} {\bibfnamefont {S.}~\bibnamefont {{Wagner}}},\ and\
  \bibinfo {author} {\bibfnamefont {E.~S.~C.}\ \bibnamefont {{Ching}}},\
  }\bibfield  {title} {\bibinfo {title} {Thermal boundary layer equation for
  turbulent {R}ayleigh-{B}{\'e}nard convection},\ }\href
  {https://doi.org/10.1103/PhysRevLett.114.114302} {\bibfield  {journal}
  {\bibinfo  {journal} {Phys. Rev. Lett.}\ }\textbf {\bibinfo {volume} {114}},\
  \bibinfo {pages} {114302} (\bibinfo {year} {2015})}\BibitemShut {NoStop}%
\bibitem [{\citenamefont {Verzicco}\ and\ \citenamefont
  {Camussi}(1999)}]{Verzicco:JFM1999}%
  \BibitemOpen
  \bibfield  {author} {\bibinfo {author} {\bibfnamefont {R.}~\bibnamefont
  {Verzicco}}\ and\ \bibinfo {author} {\bibfnamefont {R.}~\bibnamefont
  {Camussi}},\ }\bibfield  {title} {\bibinfo {title} {{P}randtl number effects
  in convective turbulence},\ }\href
  {https://doi.org/10.1017/S0022112098003619} {\bibfield  {journal} {\bibinfo
  {journal} {J. Fluid Mech.}\ }\textbf {\bibinfo {volume} {383}},\ \bibinfo
  {pages} {55–73} (\bibinfo {year} {1999})}\BibitemShut {NoStop}%
\bibitem [{\citenamefont {{Grossmann}}\ and\ \citenamefont
  {{Lohse}}(2000)}]{Grossmann:JFM2000}%
  \BibitemOpen
  \bibfield  {author} {\bibinfo {author} {\bibfnamefont {S.}~\bibnamefont
  {{Grossmann}}}\ and\ \bibinfo {author} {\bibfnamefont {D.}~\bibnamefont
  {{Lohse}}},\ }\bibfield  {title} {\bibinfo {title} {Scaling in thermal
  convection: a unifying theory},\ }\href
  {https://doi.org/10.1017/S0022112099007545} {\bibfield  {journal} {\bibinfo
  {journal} {J. Fluid Mech.}\ }\textbf {\bibinfo {volume} {407}},\ \bibinfo
  {pages} {27–56} (\bibinfo {year} {2000})}\BibitemShut {NoStop}%
\bibitem [{\citenamefont {Yang}\ \emph {et~al.}(2021)\citenamefont {Yang},
  \citenamefont {Zhang}, \citenamefont {Jin}, \citenamefont {Dong},
  \citenamefont {Wang},\ and\ \citenamefont {Zhou}}]{Yang:JFM2021}%
  \BibitemOpen
  \bibfield  {author} {\bibinfo {author} {\bibfnamefont {J.-L.}\ \bibnamefont
  {Yang}}, \bibinfo {author} {\bibfnamefont {Y.-Z.}\ \bibnamefont {Zhang}},
  \bibinfo {author} {\bibfnamefont {T.-c.}\ \bibnamefont {Jin}}, \bibinfo
  {author} {\bibfnamefont {Y.-H.}\ \bibnamefont {Dong}}, \bibinfo {author}
  {\bibfnamefont {B.-F.}\ \bibnamefont {Wang}},\ and\ \bibinfo {author}
  {\bibfnamefont {Q.}~\bibnamefont {Zhou}},\ }\bibfield  {title} {\bibinfo
  {title} {The ${P}r$-dependence of the critical roughness height in
  two-dimensional turbulent {R}ayleigh--{B}\'enard convection},\ }\href
  {https://doi.org/10.1017/jfm.2020.1091} {\bibfield  {journal} {\bibinfo
  {journal} {J. Fluid Mech.}\ }\textbf {\bibinfo {volume} {911}},\ \bibinfo
  {pages} {A52} (\bibinfo {year} {2021})}\BibitemShut {NoStop}%
\bibitem [{\citenamefont {Li}\ \emph {et~al.}(2021)\citenamefont {Li},
  \citenamefont {He}, \citenamefont {Tian}, \citenamefont {Hao},\ and\
  \citenamefont {Huang}}]{Li:JFM2021}%
  \BibitemOpen
  \bibfield  {author} {\bibinfo {author} {\bibfnamefont {X.-M.}\ \bibnamefont
  {Li}}, \bibinfo {author} {\bibfnamefont {J.-D.}\ \bibnamefont {He}}, \bibinfo
  {author} {\bibfnamefont {Y.}~\bibnamefont {Tian}}, \bibinfo {author}
  {\bibfnamefont {P.}~\bibnamefont {Hao}},\ and\ \bibinfo {author}
  {\bibfnamefont {S.-D.}\ \bibnamefont {Huang}},\ }\bibfield  {title} {\bibinfo
  {title} {Effects of {P}randtl number in quasi-two-dimensional
  {R}ayleigh--{B}\'enard convection},\ }\href
  {https://doi.org/10.1017/jfm.2021.21} {\bibfield  {journal} {\bibinfo
  {journal} {J. Fluid Mech.}\ }\textbf {\bibinfo {volume} {915}},\ \bibinfo
  {pages} {A60} (\bibinfo {year} {2021})}\BibitemShut {NoStop}%
\bibitem [{\citenamefont {Hanasoge}\ \emph {et~al.}(2012)\citenamefont
  {Hanasoge}, \citenamefont {Duvall},\ and\ \citenamefont
  {Sreenivasan}}]{Hanasoge:PNAS2012}%
  \BibitemOpen
  \bibfield  {author} {\bibinfo {author} {\bibfnamefont {S.~M.}\ \bibnamefont
  {Hanasoge}}, \bibinfo {author} {\bibfnamefont {T.~L.}\ \bibnamefont
  {Duvall}},\ and\ \bibinfo {author} {\bibfnamefont {K.~R.}\ \bibnamefont
  {Sreenivasan}},\ }\bibfield  {title} {\bibinfo {title} {Anomalously weak
  solar convection},\ }\href {https://doi.org/10.1073/pnas.1206570109}
  {\bibfield  {journal} {\bibinfo  {journal} {Proceedings of the National
  Academy of Sciences}\ }\textbf {\bibinfo {volume} {109}},\ \bibinfo {pages}
  {11928} (\bibinfo {year} {2012})},\ \Eprint
  {https://arxiv.org/abs/https://www.pnas.org/content/109/30/11928.full.pdf}
  {https://www.pnas.org/content/109/30/11928.full.pdf} \BibitemShut {NoStop}%
\bibitem [{\citenamefont {Featherstone}\ and\ \citenamefont
  {Hindman}(2016)}]{Featherstone:APJL2016}%
  \BibitemOpen
  \bibfield  {author} {\bibinfo {author} {\bibfnamefont {N.~A.}\ \bibnamefont
  {Featherstone}}\ and\ \bibinfo {author} {\bibfnamefont {B.~W.}\ \bibnamefont
  {Hindman}},\ }\bibfield  {title} {\bibinfo {title} {The spectral amplitude of
  stellar convection and its scaling in the high-{R}ayleigh-number regime},\
  }\href {https://doi.org/10.3847/0004-637x/818/1/32} {\bibfield  {journal}
  {\bibinfo  {journal} {Astrophys. J. Lett.}\ }\textbf {\bibinfo {volume}
  {818}},\ \bibinfo {pages} {32} (\bibinfo {year} {2016})}\BibitemShut
  {NoStop}%
\bibitem [{\citenamefont {Karak}\ \emph {et~al.}(2018)\citenamefont {Karak},
  \citenamefont {Miesch},\ and\ \citenamefont {Bekki}}]{Karak:POF2018}%
  \BibitemOpen
  \bibfield  {author} {\bibinfo {author} {\bibfnamefont {B.~B.}\ \bibnamefont
  {Karak}}, \bibinfo {author} {\bibfnamefont {M.}~\bibnamefont {Miesch}},\ and\
  \bibinfo {author} {\bibfnamefont {Y.}~\bibnamefont {Bekki}},\ }\bibfield
  {title} {\bibinfo {title} {Consequences of high effective {P}randtl number on
  solar differential rotation and convective velocity},\ }\href
  {https://doi.org/10.1063/1.5022034} {\bibfield  {journal} {\bibinfo
  {journal} {Phys. Fluids}\ }\textbf {\bibinfo {volume} {30}},\ \bibinfo
  {pages} {046602} (\bibinfo {year} {2018})},\ \Eprint
  {https://arxiv.org/abs/https://doi.org/10.1063/1.5022034}
  {https://doi.org/10.1063/1.5022034} \BibitemShut {NoStop}%
\bibitem [{\citenamefont {Vasil}\ \emph {et~al.}(2021)\citenamefont {Vasil},
  \citenamefont {Julien},\ and\ \citenamefont {Featherstone}}]{Vasil:PNAS2021}%
  \BibitemOpen
  \bibfield  {author} {\bibinfo {author} {\bibfnamefont {G.~M.}\ \bibnamefont
  {Vasil}}, \bibinfo {author} {\bibfnamefont {K.}~\bibnamefont {Julien}},\ and\
  \bibinfo {author} {\bibfnamefont {N.~A.}\ \bibnamefont {Featherstone}},\
  }\bibfield  {title} {\bibinfo {title} {Rotation suppresses giant-scale solar
  convection},\ }\bibfield  {journal} {\bibinfo  {journal} {Proc. Natl. Acad.
  Sci. USA}\ }\textbf {\bibinfo {volume} {118}},\ \href
  {https://doi.org/10.1073/pnas.2022518118} {10.1073/pnas.2022518118} (\bibinfo
  {year} {2021}),\ \Eprint
  {https://arxiv.org/abs/https://www.pnas.org/content/118/31/e2022518118.full.pdf}
  {https://www.pnas.org/content/118/31/e2022518118.full.pdf} \BibitemShut
  {NoStop}%
\end{thebibliography}
\end{document}